\documentclass[fleqn,usenatbib]{mnras}
\usepackage{mathptmx,subfigure,relsize}
\usepackage[T1]{fontenc}
\DeclareRobustCommand{\VAN}[3]{#2}
\let\VANthebibliography\thebibliography
\def\thebibliography{\DeclareRobustCommand{\VAN}[3]{##3}\VANthebibliography}

\usepackage{graphicx}
\graphicspath{{./plots/}}
\usepackage{amsmath}
\usepackage{cleveref}
\usepackage{upgreek}
\usepackage{tipa}
\usepackage{amssymb}
\usepackage[T1]{fontenc}

\newcommand*{\eLR}{${\rm Ref-L050N0752}\,\,$}
\newcommand*{\LR}{${\rm LoResDM}$}
\newcommand*{\HR}{${\rm HiResDM}$}
\newcommand{\subfind}{\textsc{subfind}{ }}
\newcommand{\eagle}{\textsc{eagle}}

\newcommand{\gadget}{\textsc{Gadget3}{ }}
\def\ltsima{$\; \buildrel < \over \sim \;$}
\def\simlt{\lower.5ex\hbox{\ltsima}}
\def\gtsima{$\; \buildrel > \over \sim \;$}
\def\simgt{\lower.5ex\hbox{\gtsima}}
\def\la{$\langle$}

\title[Spurious heating of simulated galaxies]{Spurious heating of stellar motions by dark matter particles in cosmological simulations of galaxy formation}

\author[Ludlow et al.] {\parbox{18cm}{
    Aaron D. Ludlow$^{1,\star}$,
    S. Michael Fall$^{2}$,
    Matthew J. Wilkinson$^{1}$, 
    Joop Schaye$^{3}$,
    Danail Obreschkow$^{1}$\\
  }\vspace{0.3cm}\\
  $^{1}${International Centre for Radio Astronomy Research, University of Western Australia, 35 Stirling Highway, Crawley,}\\
  {Western Australia, 6009, Australia}\\
  $^{2}${Department of Physics and Astronomy, Johns Hopkins University, 3400 N. Charles St., Baltimore, MD, 21218, USA}\\
  $^{3}${Leiden Observatory, Leiden University, PO Box 9513, 2300 RA Leiden, the Netherlands}\\
}

\date{Accepted XXX. Received YYY; in original form ZZZ}

\pubyear{2021}

\hypersetup{draft}
\begin{document}
\label{firstpage}
\pagerange{\pageref{firstpage}--\pageref{lastpage}}
\maketitle

\begin{abstract}  
  We use two cosmological simulations to study the impact of spurious heating of stellar motions within simulated
  galaxies by dark matter (DM) particles. The simulations share the same numerical and subgrid parameters,
  but one used a factor of 7 more DM particles. Many galaxy properties are unaffected by
  spurious heating, including their masses, star formation histories, and the spatial distribution of
  their gaseous baryons. The distribution and kinematics of stellar and DM particles, however,
  are affected. Below a resolution-dependent virial mass, $M_{200}^{\rm spur}$, galaxies have higher
  characteristic velocities, larger sizes, and more angular momentum in the simulation with lower DM
  mass resolution; haloes have higher central densities and lower velocity dispersions. Above
  $M_{200}^{\rm spur}$, galaxies and haloes have similar properties in both runs. The differences arise
  due to spurious heating, which transfers energy from DM to stellar particles, causing galaxies to heat
  up and haloes to cool down. The value of $M_{200}^{\rm spur}$ can be derived from an empirical disc heating
  model, and coincides with the mass below which the predicted {\em spurious} velocity dispersion exceeds the
  {\em measured} velocity dispersion of simulated galaxies. We predict that galaxies in the $100^3\, {\rm Mpc}^3$
  \eagle\, run and IllustrisTNG-100 are robust to spurious collisional effects at their half-mass radii
  provided $M_{200}^{\rm spur}\approx 10^{11.7}{\rm M_\odot}$; for the
  $25^3\, {\rm Mpc}^3$ \eagle\, run and IllustrisTNG-50, we predict $M_{200}^{\rm spur}\approx 10^{11}{\rm M_\odot}$.
  Suppressing spurious heating at smaller/larger radii, or for older/younger stellar populations, requires
  haloes to be resolved with more/fewer DM particles.
\end{abstract}


\renewcommand{\thefootnote}{\fnsymbol{footnote}}
\footnotetext[1]{E-mail: aaron.ludlow@icrar.org}

\section{Introduction}
\label{SecIntro}

Relaxation driven by scattering between stars or dark matter (DM) particles is believed to be negligible for
most galaxies. Simulations of galaxy formation aim to emulate this collisionless behaviour, and
although they continue to improve, most do not model individual systems with particle numbers that
are even close to the number of stars, let alone DM particles, in real galaxies; they instead employ fewer, more
massive ones. Simulated galaxies and their DM haloes are therefore susceptible to spurious evolution due to the
finite number of particles with which they are resolved, which gives rise to incoherent 
fluctuations in their gravitational potential that deflect individual particle trajectories
\citep[e.g.][]{LO1985,Hernquist1990b,Sellwood2013,Sellwood2015}.
This gravitational scattering process (that we refer to as ``spurious heating'')
can be detrimental to the structure and kinematics of simulated galaxies.

Using a suite of idealised simulations of isolated
galactic discs, \citet{Ludlow2021} and \citet{Wilkinson2023} showed that spurious
heating randomises the orbits of stellar particles, thus increasing the velocity dispersion and 
thickness of discs, and affecting the locations of galaxies on a number
of fundamental scaling relations. They provided a simple empirical model that 
quantifies the effects of spurious heating on the velocity dispersion and scale heights of stellar disc particles.
For simulations that adopt DM particle masses of $m_{\rm DM}\gtrsim 10^6\,{\rm M_\odot}$,
such as \eagle~\citep{Schaye2015,Crain2015} and IllustrisTNG-100 \citep{Pillepich2018},
their model predicts that spurious heating can be significant in haloes with masses below about
$10^{12}\,{\rm M_\odot}$, suggesting that many galaxies of interest in large-volume cosmological
simulations will suffer ill effects.

Although spurious heating is critically important for simulations of idealised galactic discs, the consequences for
galaxies in cosmological simulations is less clear. Idealised simulations, for example, neglect a variety of important
physical processes that also give rise to disk heating: the scattering of stellar orbits by spiral arms, 
globular clusters, or giant molecular clouds, for example, or dynamical heating driven by accretion or close encounters with
nearby satellite galaxies. Idealised simulations also conserve mass and typically disregard the time-dependent growth
of the DM halo as well as the formation of new stars in a thin, rotationally supported disk. When acting in concert,
these processes will either exacerbate or mitigate the effects of spurious heating in cosmological runs. 

Spurious heating, however, {\em is present} in cosmological simulations of galaxy formation. 
\citet[][see also \citealt{Revaz2018}]{Ludlow2019} showed that galaxies in the intermediate resolution
\eagle~simulation (for which $m_{\rm DM}=9.7\times 10^6{\rm M_\odot}$) with
stellar masses $M_\star\lesssim 10^{10}\,{\rm M_\odot}$ are subject to spurious size growth due to collisional
heating. But what other galaxy properties inferred from cosmological simulations are affected by spurious heating?
Do the insights obtained from idealised simulations apply to galaxies formed in cosmological ones? And how, if at all,
can we distinguish spurious results from robust ones? This paper addresses these questions.

Our paper is structured as follows. In Section~\ref{SecSims} we describe our simulations and analysis techniques.
Our main results are presented in Section~\ref{sec:galpop}, where we highlight several galaxy properties that
{\em are not} affected by spurious heating; and in Section~\ref{sec:scaling}, where we identify and compare properties of
galaxies that {\em are} affected by spurious heating. In Section~\ref{sec:imp} we discuss the implications of our results 
for resolving the inner regions of galaxies and haloes, and provide a simple convergence criterion that can be used to separate
galaxies whose
properties are robust to spurious heating from those whose properties are not. We summarise our results in Section~\ref{sec:summary}.

\section{Numerical simulations and Analysis}
\label{SecSims}

\begin{center}
  \begin{table*}
    \caption{Numerical aspects of our simulations. From left to right: simulation name, box side length ($L_{\rm box}$), the total number
      of dark matter particles and the initial number of baryonic particles ($N_{\rm DM}$ and $N_{\rm gas}$, respectively), dark matter and
      (primordial) baryonic particle masses ($m_{\rm DM}$ and $m_{\rm gas}$, respectively), and dark matter-to-baryonic particle mass ratio
      ($\mu\equiv m_{\rm DM}/m_{\rm gas}$). $N_{\rm cent}$ and $N_{\rm sub}$ are the total number of central haloes and satellite
      subhaloes, respectively, with masses greater than $100\times m_{\rm DM}$. Similarly, $N_{\rm gal}$ and $N_{\rm sat}$ are the
      number of central and satellites galaxies, respectively, with stellar masses greater than $10\times m_{\rm gas}$.}
    \begin{tabular}{l l c r c c c c c c c c c c c}\hline \hline
      & ${\rm Sim.\,\,Name}$ & $L_{\rm box}$ & $N_{\rm DM}$ & $N_{\rm gas}$ & $m_{\rm DM}$           & $m_{\rm gas}$          & $\mu$ & $N_{\rm cent}$ & $N_{\rm sub}$ & $N_{\rm gal}$ & $N_{\rm sat}$ &\\
      &                      & ${\rm [Mpc]}$   &                & & $[10^5 {\rm M_\odot}]$ &  $[10^5{\rm M_\odot}]$ &  $m_{\rm DM}/m_{\rm gas}$ & $\geq 100\, m_{\rm DM}$ & $\geq 100\, m_{\rm DM}$ & $\geq 10\, m_{\rm gas}$ & $\geq 10\, m_{\rm gas}$ &\\\hline
      & \LR &  50             &   $752^3$      &   $752^3$       &     97.0               &   18.1                 &       5.36                & \multicolumn{1}{|r|}{70,338} & \multicolumn{1}{|r|}{22,141} & 7,932 & 5,551 &\\
      & \HR &  50             &$7\times 752^3$ &   $752^3$       &     13.9               &   18.1                 &       0.77                & \multicolumn{1}{|r|}{445,665}& \multicolumn{1}{|r|}{128,243} & 7,483& 6,139&\\\hline
    \end{tabular}
    \label{TabSimParam}
  \end{table*}
\end{center}

\subsection{The simulations}
\label{ssec:sims}

We analyse the $z=0$ outputs of two cosmological simulations
of galaxy formation that differ only in the resolution of their
DM components, but keep the baryon mass resolution, force softening, and subgrid physics fixed.
One simulation is \eLR of the \eagle~ project (see Table 2 of \citealt{Schaye2015}),
which follows the evolution of cosmic structure in a comoving cube with edge lengths $L_{\rm box}=50\,{\rm Mpc}$ 
initially using $N_{\rm DM}=N_{\rm gas}=752^3$ particles of both DM and baryons. The second simulation uses the same box
size and number of baryonic particles, but to study the effects of higher DM mass resolution we split each DM particle
equally into 7 lower-mass ones. Throughout the paper we refer to these runs as \LR\,and \HR, respectively.\footnote{A similar
  pair of simulations was carried out by \citet{Richings2021} but in a smaller volume, $L_{\rm box}=25\,{\rm Mpc}$. These
  were the same runs used by \citet{Ludlow2019} to study the impact of spurious heating on simulated galaxy sizes. We refer
  to those papers for additional information.}

Both simulations were run with the same version of \gadget \citep{Springel2005b,Schaye2015} that was used for \eagle,
and employed the same numerical parameters for hydrodynamics and gravitational force integration.
We adopted the ``Reference'' subgrid models and parameters for cooling and star formation, and for feedback from stars
and active galactic nuclei \citep[see][for details]{Schaye2015}; the parameters were not recalibrated for \HR\, to
accommodate any effects of its higher DM mass resolution.

We adopt cosmological parameters advocated by the \citet{Planck2014}.
The initial\footnote{Because of star formation and mass transfer between stellar and gas particles the masses of
baryonic particles may evolve with time.} baryonic particle mass is therefore $m_{\rm gas}=1.81\times 10^6\,{\rm M_\odot}$
for both runs, and the DM particle masses are $m_{\rm DM}=9.70\times 10^6\,{\rm M_\odot}$ for
\LR\footnote{This is the same DM particle mass that was used for the 100 cubic Mpc flagship
run of the \eagle\, project, referred to as L100N1504 in Table 2 of \citet{Schaye2015}.}
and $m_{\rm DM}=1.39\times 10^6\,{\rm M_\odot}$ for \HR.
We will sometimes quote the DM-to-baryonic particle mass ratio, i.e. $\mu\equiv m_{\rm DM}/m_{\rm gas}$, which
is initially $\mu \approx 5.36$ for \LR\, and $\approx 0.77$ for \HR.

The gravitational softening lengths are the same for both runs for all particle species; they
were chosen to be a fixed fraction of the Lagrangian baryonic inter-particle separation,
$l_{\rm bar}=L_{\rm box}/N_{\rm gas}^{1/3}$. For redshifts $z>2.8$ the value is $\epsilon/l_{\rm bar}=0.04$ in comoving units
and at lower redshifts it is $\epsilon/l_{\rm bar}=0.011$ in physical units; at $z\leq 2.8$, this corresponds to a physical
softening length $\epsilon=700\,{\rm pc}$. Inter-particle forces are exactly Newtonian above the {\em spline} softening length,
$\epsilon_{\rm sp}=2.8\times\epsilon$.

\subsection{Halo identification and matching}
\label{ssec:haloes}

We use \subfind to identify gravitationally bound DM haloes and their associated galaxies
\citep{Springel2001b}. \subfind identifies friends-of-friends
(FoF) DM haloes and divides them into self-bound ``subhaloes''.
We refer to the most massive subhalo of each FoF group as the {\em central} subhalo; the rest are satellite subhaloes. 
\subfind also associates baryonic particles to each central and satellite subhalo -- we refer to them 
as central and satellite {\em galaxies}, respectively. The centres of haloes, subhaloes, and galaxies
coincide with their DM particle that has the lowest gravitational potential energy.

On mutually-resolved scales, both runs employed the same initial Fourier modes for density fluctuations allowing individual
haloes in \LR\, to be matched to their counterparts in \HR. But because the total number of DM particles differ,
matching haloes by their DM particle IDs is not possible. We instead adopt a different approach, as follows.

For each central halo in \HR\, at $z=0$, we select 50 DM particles with the lowest
potential energies and determine their Lagrangian coordinates in the initial conditions (ICs) of the simulation. 
We next locate the 50 nearest DM particles in the ICs of \LR, and determine the $z=0$ halo (if any) to which
the majority of them belong; we accept a halo in \LR\,as a potential match to the one in \HR\,if it contains at least 25 of them.
We then repeat the procedure in reverse, using haloes in \LR\,as the starting point. We note that 
$\gtrsim 96$ per cent of centrals with total masses $\gtrsim 10^{10}\,{\rm M_\odot}$ were bijectively matched between
the two runs.

Table~\ref{TabSimParam} lists some important numerical aspects of the simulations, as well as some basic properties of
the halo and galaxy populations.

\subsection{Analysis}
\label{ssec:anal}

We define the virial masses\footnote{The virial mass, $M_{200}$, is the total mass of all particle species enclosed
by a sphere of radius $r_{200}$ (measured relative to the halo centre) that encloses a mean density equal to
$200\times \rho_{\rm crit}(z)$, where $\rho_{\rm crit}=3 H^2(z)/8 \pi G$ is the critical density for closure at
redshift $z$, $G$ is the gravitational constant, and $H(z)$ is the Hubble-Lema\^itre parameter. The corresponding
virial circular velocity is $V_{200}=(G\,M_{200}/r_{200})^{1/2}$.} of central galaxies (hereafter centrals) as $M_{200}$.
The masses of satellite haloes, $M_{\rm sub}$, are defined as the sum of the masses of their gravitationally bound particles (baryonic
plus dark matter). The stellar or gas mass of a central or satellite galaxy -- $M_\star$ or
$M_{\rm gas}$, respectively -- is defined as the sum of the masses of all gravitationally bound
particles of that type.\footnote{Several earlier studies based on \eagle\, defined $M_\star$ as the integrated mass
of bound stellar particles contained within a fixed physical aperture, commonly $30\,{\rm kpc}$. Adopting this definition 
changes slightly the masses of a handful of the most massive galaxies in our runs but otherwise has no impact on our results.}

In addition to masses, we also calculate a number of dynamical and structural properties of the galaxies.
We adopt a coordinate system centred on each galaxy and at rest with respect to its centre-of-mass
motion (defined as the mass-weighted mean velocity of its stellar particles); the $z$-axis
is chosen to coincide with the total (stellar) angular momentum vector of the galaxy.

Galaxy sizes are quantified by their three dimensional stellar half-mass radius, $r_{50}$ (we will use lower case ``$r$''
and capital ``$R$'' to distinguish spherical and cylindrical radii, respectively; in our coordinate system,
$R^2\equiv r^2-z^2$).

\begin{figure}
  \subfigure{\includegraphics[width=0.48\textwidth]{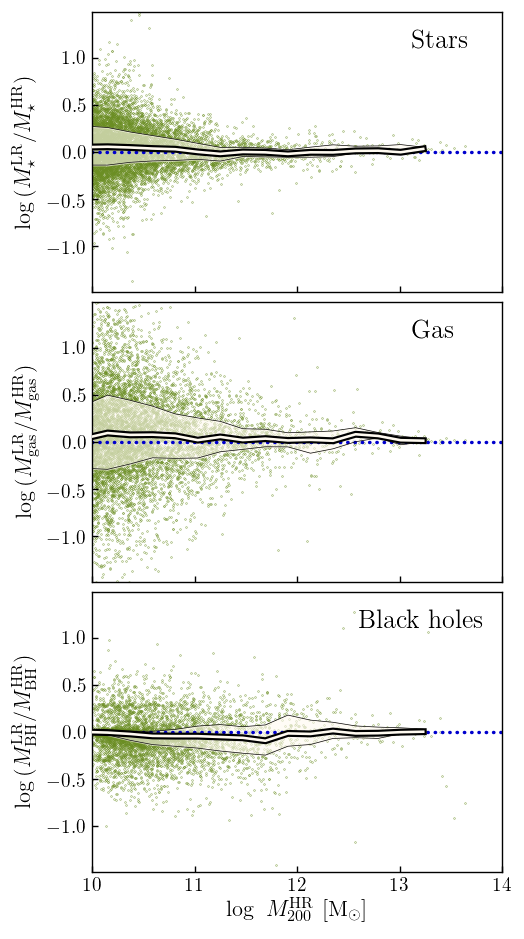}}
  \caption{Logarithmic difference between the stellar (top panel), gas (middle panel), and central supermassive
    black hole masses (lower panel) of individual central galaxies that were matched between \LR\, and \HR. 
    Results are plotted as a function of the virial mass in the \HR\, run, i.e. $M_{200}^{\rm HR}$.
    The thick lines with black boundaries show the median relations and the lightly shaded regions indicate
    the interquartile scatter. Individual galaxies are plotted using green points. 
    Each of these baryonic masses are strongly correlated, although the scatter
    in $M_\star$ and $M_{\rm gas}$ increases toward lower halo masses. 
    For BHs, the small scatter at low halo masses is due to the majority of those galaxies hosting single BHs with
    masses approximately equal to their seed mass.}
  \label{fig:mscatter}
\end{figure}

\begin{figure}
  \subfigure{\includegraphics[width=0.48\textwidth]{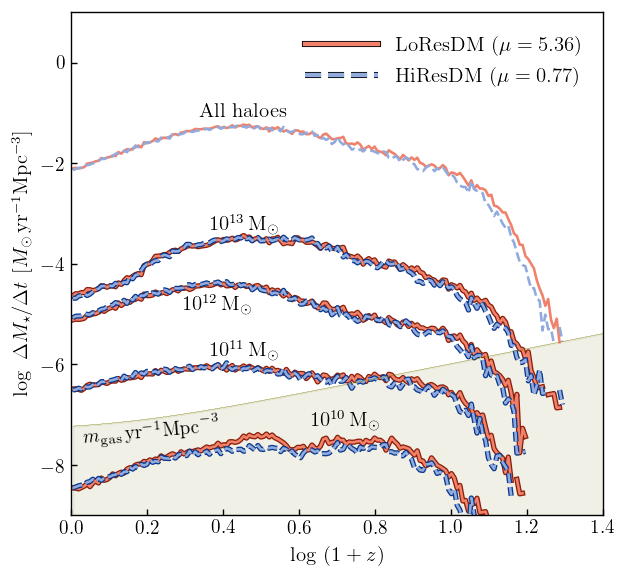}}
  \caption{Average star formation histories (SFHs) for central galaxies occupying DM haloes with different $z=0$
    virial masses (the values of $M_{200}$ are chosen to lie within $\pm 0.15$ dex of the values indicated near
    each pair of curves). Results are shown for \LR\, using solid orange lines and for \HR\, using dashed
    blue lines. The uppermost set of curves (plotted as thin lines) show the total
    star formation history in each simulation. The shaded region indicates star formation rates
    less than $m_{\rm gas} {\rm yr}^{-1} {\rm Mpc}^{-3}$.
    Present-day star formation rates as well as past star formation
    histories are, on average, unaffected by the higher DM mass resolution used for \HR. }
  \label{fig:sfh}
\end{figure}

We quantify the thickness of galactic discs using the stellar half-mass height, $z_{50}$, 
calculated in an annular aperture of central radius $R=r_{50}$ and width $\Delta\log R=0.3$. In the same aperture we also
calculate the stellar velocity dispersion, $\sigma_{\star,50}$, defined as\footnote{In Appendix~\ref{appB}, we also consider
  characteristic radii $r_f$ that enclose different stellar mass fractions $f$, namely the radii $r_{25}$ and $r_{75}$
  that enclose 25 and 75 per cent of the galactic stellar mass, respectively. The local velocity dispersion at $r_f$ will be denoted
  explicitly $\sigma_{\star,f}$.}
\begin{equation}
  \sigma_{\star,50}^2=\frac{\sum_k \, m_k v_k^2}{\sum_k \, m_k},
  \label{eq:sigma}
\end{equation}
where $v_k$ is the velocity norm of particle $k$ in the reference frame of the galaxy, $m_k$ is its mass, and the sums extend over all
particles $k$ in the annulus. Note that we do not subtract the mean azimuthal velocities of stellar particles;
$\sigma_{\star,50}^2$ is therefore equal to twice the specific kinetic energy of stellar particles within the shell. 

We construct spherically averaged circular velocity profiles due to DM, stellar, and gas particles, and define the total circular velocity
profile (squared) as
\begin{equation}
  V_c^2(r)\equiv \frac{G\, M_{\rm tot}(r)}{r}=V_{\rm DM}^2(r)+V_\star^2(r)+V_{\rm gas}^2(r),
  \label{eq:Vc}
\end{equation}
where $M_{\rm tot}(r)$ is total enclosed mass and
the subscripts indicate the matter components.
Although strictly valid only for spherical systems, we use equation~(\ref{eq:Vc}) as a proxy for the galactic 
rotation velocity, which we define as the total circular velocity at $r_{50}$: 
$V_{c,50}\equiv V_c(r_{50})$. When defined this way, late-type galaxies in \eagle\, and IllustrisTNG-100 with 
$M_\star\gtrsim 10^{10} {\rm M_\odot}$ follow a Tully-Fisher (\citeyear{TF1977}) relation that agrees well
with observations \citep{Ferrero2017}. 

Scaling laws that relate the above quantities often depend on galaxy morphology. 
We separate galaxies by morphological type 
using $\kappa_{\rm co}$, defined as the fraction of kinetic energy in stellar orbits
co-rotating with the galaxy\footnote{Specifically, we adopt the definition
$\kappa_{\rm co}=(2\,K_\star)^{-1}\sum_{j_{z,k}>0} m_k\,(j_{z,k}/R_k)^2$, where $j_{z,k}$ is the vertical component of the
specific angular momentum of particle $k$ and $K_\star$ is the total kinetic energy of the stellar particles.}
\citep{Sales2010,Correa2017}. \citet{Correa2020} showed that, for $M_{200}\gtrsim 10^{12}{\rm M_\odot}$,
the halo mass dependence of the disc and spheroid fractions in the Sloan Digital Sky Survey (Data Release 7; 
\citealt{York2000,Abazajian2009}) are reproduced by \eagle~using the thresholds $\kappa_{\rm co}\geq 0.35$ and
$\kappa_{\rm co}\leq 0.25$, respectively.
We note, however, that more direct methods of estimating galactic morphology -- based, for example, on disk-to-bulge
ratios, or dynamical decomposition of galactic structures -- may yield different boundaries between discs and spheroids
whose structural scaling relations
may differ from those presented in Section~\ref{sec:scaling}. We plan to study the impact of spurious heating
on galactic morphology in a follow-up paper; for the present paper, we adopt the $\kappa_{\rm co}$ thresholds
advocated by \citet{Correa2020} without further justification.

It will sometimes be necessary to distinguish a quantity measured in one of our simulations from the same
quantity measured in the other. We will do so using superscripts: ``${\rm HR}$'' for \HR\, and ``${\rm LR}$''
for \LR. For example, the stellar mass of a galaxy in \LR\, will sometimes be denoted $M_\star^{\rm LR}$.

\begin{figure}
  \subfigure{\includegraphics[width=0.48\textwidth]{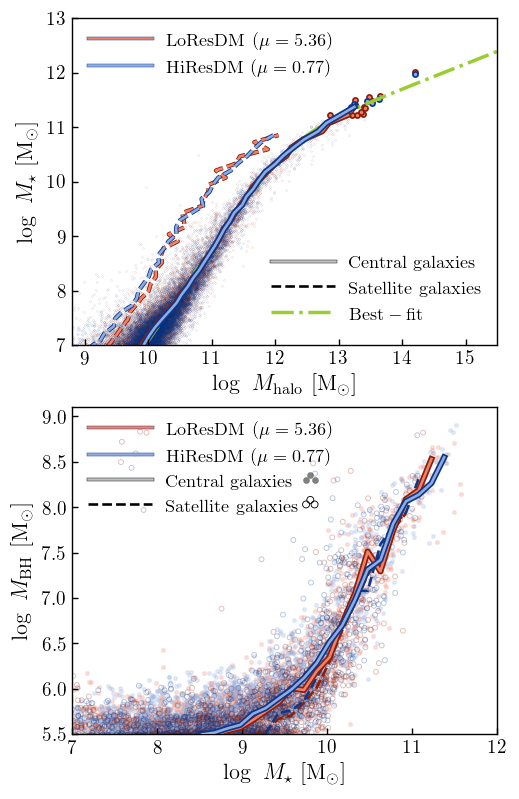}}
  \caption{Upper panel: Median stellar-to-halo mass relation for central (solid lines) and satellite galaxies
    (dashed lines). For centrals, halo masses are defined as $M_{200}$; for satellites, we use the total
    gravitationally bound mass, $M_{\rm sub}$, identified by \textsc{subfind}. 
    Individual central galaxies are shown as dots, or as circles in bins populated by
    fewer than 10 systems. As in Fig.~\ref{fig:sfh}, orange lines show the median relations for \LR\,and blue lines for \HR. The
    green dashed line is the best-fitting double power-law to the $M_\star - M_{\rm halo}$ relation of centrals.
    Lower panel: Black hole mass-stellar mass relation for central and satellite galaxies. Median relations and
    individual systems are shown as solid lines and filled circles, respectively, for centrals, and as dashed lines
    and open circles, respectively, for satellites.}
  \label{fig:ms_mbh_mh}
\end{figure}

\section{Galaxy properties that are not significantly affected by spurious heating}
\label{sec:galpop}

In this section we present several important properties of the simulated galaxy population that are {\em not} significantly
affected by spurious heating; In Section~\ref{sec:scaling} we present a few properties that {\em are} significantly affected.

\subsection{Stellar, gas, and central supermassive black hole masses of central galaxies}
\label{ssec:masses}

Fig.~\ref{fig:mscatter} plots the logarithmic difference between the stellar masses (top panel), gas masses (middle
panel), and central black hole masses (bottom panel) of matched centrals as a function of $M_{200}^{\rm HR}$, the
halo virial mass in \HR.\footnote{We find that the virial masses of matched galaxies used in our analysis are in excellent
agreement, typically within a few per cent, and with no systematic bias. Using $M_{200}^{\rm HR}$, $M_{200}^{\rm LR}$,
or their average as the independent variable in Fig.~\ref{fig:mscatter} does not
affect the results.} Green symbols show individual galaxies and the thick lines with black boundaries are the median
relations. Note that all three baryonic masses are in good agreement, even down to the lowest
values of $M_{200}^{\rm HR}$ plotted ($10^{10}{\rm M_\odot}$).

The lightly shaded regions in each panel indicate the interquartile range (IQR). For $M_\star$ and $M_{\rm gas}$,
the scatter decreases with increasing halo mass (although it is larger for $M_{\rm gas}$ than for $M_\star$ at all masses).
For example, galaxies with $M_\star\geq 10^{10}{\rm M_\odot}$ have an interquartile scatter about the one-to-one logarithmic
relation of ${\rm IQR}=0.09$ for $M_\star$ and ${\rm IQR}=0.16$ for $M_{\rm gas}$; for $M_\star< 10^{10}\,{\rm M_\odot}$, the
scatter increases to ${\rm IQR}=0.29$ for $M_\star$ and ${\rm IQR}=0.54$ for $M_{\rm gas}$.

The scatter cannot be attributed to Poisson noise alone, for which the relative mass
difference is expected to be $\sim 1/\sqrt{N_i}$, where $N_i$ is the number of particles of species $i$. As discussed by
\citet{Borrow2022b}, it is partially due to stochastic differences in the star formation histories of
individual galaxies that arise due to the star formation and feedback models adopted for \eagle\,
\citep[see also][]{Genel2019,Keller2019}, but the spurious collisional effects discussed in this paper may also contribute.

The mass-dependence of the scatter in BH masses differs from that for $M_\star$ and $M_{\rm gas}$. In this case,
the scatter about the one-to-one logarithmic relation at $M_\star< 10^{10}\,{\rm M_\odot}$ (${\rm IQR}=0.02$)
is considerably lower than it is at $M_\star\geq 10^{10}\,{\rm M_\odot}$ (${\rm IQR}=0.13$).
At sufficiently low masses, $M_{200}\lesssim 10^{10.4}\,{\rm M_\odot}$, most galaxies in both runs
host only one BH particle whose mass is roughly equal to the BH seed mass (i.e. $M_{\rm BH}=10^{5.5}\,{\rm M_\odot}$) and
the interquartile scatter drops to zero. The remaining scatter at these masses is due to small differences
in the BH accretion rates in low-mass haloes, or because galaxies in one simulation contain central BHs that have undergone
one or a few mergers with other seed-mass BHs but their counterparts in the other simulation have not. 

\subsection{Star formation histories and star formation rates}

The strong correlation between the stellar masses of individual galaxies shown in Fig.~\ref{fig:mscatter}
results from their similar star formation histories (SFHs).
We show this explicitly in Fig.~\ref{fig:sfh}, where we plot the average SFHs of centrals occupying
DM haloes with present-day
virial masses ranging from $M_{200}=10^{10}\,{\rm M_\odot}$ to $10^{13}\,{\rm M_\odot}$ (the corresponding number of
DM particles within $r_{200}$ range from 
$N_{200}\approx 10^3$ to $\approx 10^6$ in \LR, and a factor of 7 more DM particles in \HR); for comparison,
the total star formation rate densities averaged over all haloes are shown as thin lines.
Results from \LR\, are plotted using solid orange curves and those from \HR\, as dashed blue curves (a
colour convention used for the majority of plots that follow).  

The average SFHs of galaxies are
remarkably similar for \LR\, and \HR\, at all redshifts and for all $z=0$ halo masses. This
suggests that cosmic star formation is
largely unaffected by spurious heating, at least for the DM particle masses and the DM-to-stellar particle mass ratios
studied here (i.e. $\mu\approx 5.36$ and 0.77). Good agreement in
the specific star formation rates of galaxies at redshifts $z\geq 0$ is implicit in these results. 

We have verified that the mass fraction, as well as the spatial and density distribution of star
forming gas particles inside individual haloes are also well-converged. 
\citet{Steinmetz1997} showed that discreteness effects give rise to spurious heating of gas particles 
by DM halo particles that can counteract the effects of radiative cooling in poorly resolved haloes, or in the
central regions of massive ones. However, for the typical baryon fractions of $\Lambda$CDM haloes and for standard cooling
functions, this type of heating only becomes
important in simulations that adopt DM particle masses $m_{\rm DM}\gtrsim 10^9{\rm M_\odot}$,
which is much larger than the values used for our runs.

\subsection{The stellar-halo and stellar-black hole mass relations}

\begin{figure}
  \includegraphics[width=0.48\textwidth]{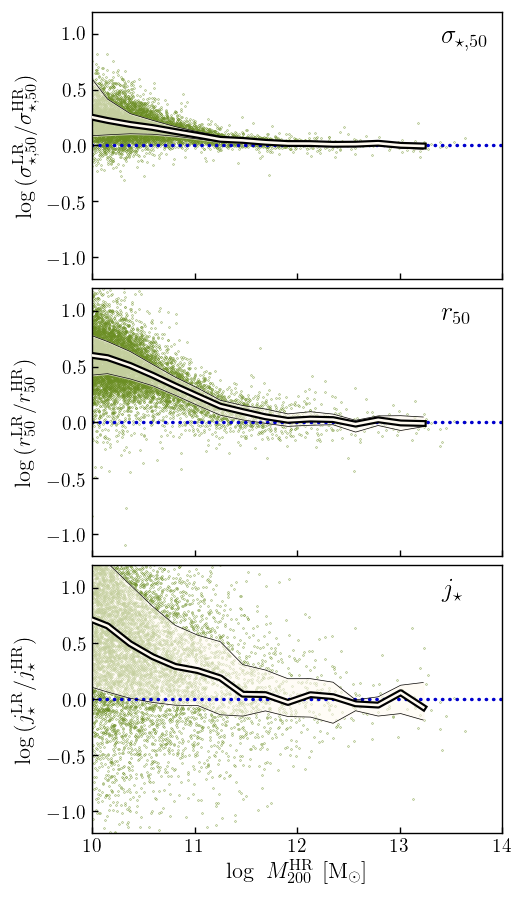}
  \caption{Logarithmic difference between the characteristic velocity dispersion ($\sigma_{\star,50}$; top panel),
    half-mass radius ($r_{50}$; middle panel), and specific angular momentum ($j_\star$; bottom panel)
    of central galaxies that were cross-matched between our simulations; results are plotted
    as a function of $M_{200}^{\rm HR}$. As in Fig.~\ref{fig:mscatter}, individual galaxies are shown using green points,
    the median relations using thick grey lines (bounded by black lines), and the interquartile
    scatter using lightly shaded regions. Above a halo mass of $\approx {\rm a\,\,few}\times 10^{11}{\rm M_\odot}$, 
    these structural and kinematic quantities are largely unbiased and exhibit small scatter between the runs. Below
    that mass, however, all three quantities exhibit increased scatter and become increasingly biased toward higher values
    in \LR, suggesting that spurious heating leads to an increase in the velocity dispersion, size, and
    angular momentum of low-mass galaxies. }
  \label{fig:dscatter}
\end{figure}

The good agreement between the stellar and black hole masses of central galaxies apparent in Fig.~\ref{fig:mscatter} has
implications for a number of important statistics of the galaxy population, e.g. 
the stellar-halo and stellar-black hole mass relations. Because of their importance as diagnostics for
subgrid model calibration, we plot these relations explicitly in Fig.~\ref{fig:ms_mbh_mh}, using solid and dashed lines
to differentiate the median relations obtained for central and satellite galaxies, respectively
(note that $M_{200}$ and $M_{\rm sub}$ define the total halo masses
of central and satellites, respectively). 

The median $M_\star - M_{\rm halo}$ relations (upper panel)
closely overlap at all stellar masses plotted (i.e. $M_\star\gtrsim 10^7\,{\rm M_\odot}$,
which is roughly equivalent to the mass of 5 primordial baryonic particles). The horizontal offset between the
relations obtained for centrals and satellites is largely due to tidal stripping, which reduces the halo
masses of satellite galaxies more readily than their stellar masses \citep[e.g.][]{Fattahi2018}. The dots, which show individual central
galaxies, suggest that the scatter in the relations are also similar, not surprising given the strong correlation
between the stellar masses of individual galaxies plotted in Fig.~\ref{fig:mscatter}
(for clarity, individual satellites are not shown). The dashed green line shows the best-fitting double power-law\footnote{The
double power-law model is $M_\star/M_{\rm halo}=A\,[(M_{\rm halo}/M_0)^{-\beta}+(M_{\rm halo}/M_0)^\gamma]^{-1}$,
with parameter values $A=0.0702$, $M_0=10^{11.59}\,{\rm M_\odot}$, $\beta=1.376$, and $\gamma=0.608$. \label{fn_fit}} to the
$M_\star - M_{\rm 200}$ relation of centrals.

In the bottom panel we plot the stellar mass-black hole mass relations obtained from our simulations.
The solid circles show individual central galaxies and the open circles individual satellites. As with the
stellar-to-halo mass relation, the stellar-black hole mass relations are in excellent agreement.

As a corollary of these results, the galaxy stellar mass functions, and the mass functions of
gaseous baryons, central black holes, and dark matter haloes are also well converged in our runs,
typically agreeing to better than $10$ per cent at all mass scales plotted in the figures above.
This applies to the full galaxy population, but also separately to central and satellite galaxies
(see Appendix~\ref{appA}).

\section{Galaxy properties that are affected by spurious heating}
\label{sec:scaling}

The results presented in the previous section suggest that the star formation rates 
and baryonic masses of cross-matched galaxies are similar in our runs, with no systematic
differences between them. We have also verified that the spatial and density distributions of the star forming gas
particles that they contain are also similar. Any differences in the structure or kinematics of stellar
particles within haloes long after their formation must therefore arise from subsequent processes. This
justifies using our simulations to study the effects of spurious heating on stellar motions in simulated
galaxies. We explore this below.

\subsection{Characteristic stellar velocities, sizes and angular momenta}
\label{ssec:spur}

Fig.~\ref{fig:dscatter} shows the logarithmic difference between the two simulations for the
characteristic velocity dispersion ($\sigma_{\star,50}$;
top panel), stellar half-mass radius ($r_{50}$; middle column), and specific angular momentum ($j_\star$; bottom panel) of
matched centrals. We use plotting conventions familiar from Fig.~\ref{fig:mscatter}.
This figure highlights one of our main results, namely that the kinematic and structural properties of
galaxies -- particularly low-mass ones -- are affected by spurious heating.\footnote{This implicitly assumes
that, at fixed $M_{200}$, galaxies in \HR\, are substantially less affected by spurious heating than those
in \LR.} At a halo mass of $M_{200}\approx 10^{11}\,{\rm M_\odot}$, for example, $\sigma_{\star,50}$ is on
average $\approx 30$ per cent higher for galaxies in \LR\,than for those in \HR, despite them having similar
baryonic and total masses (this bias is larger than the interquartile halo-to-halo scatter at this mass,
shown as lightly shaded regions in each panel of Fig.~\ref{fig:mscatter}). Although this kinematic bias
diminishes with increasing mass, at the mass resolution of \eagle\, it is noticeable for 
$M_{200}\lesssim 10^{12}\,{\rm M_\odot}$ (corresponding to $M_\star\lesssim 10^{10}\,{\rm M_\odot}$); for
$M_{200}\gtrsim 10^{12}\,{\rm M_\odot}$, however, the
bias has diminished considerably (for quantities measured at $r_{50}$, but the bias is present at smaller radii; see
Appendix~\ref{sec:radii}).

As discussed by \citet{Ludlow2019}, the increased velocity dispersion that arises due to spurious heating
leads to artificial size growth.
This can be seen in the middle panel of Fig.~\ref{fig:dscatter}.
At $M_{200}\approx 10^{11}\,{\rm M_\odot}$, the average value of $r_{50}$ is nearly a factor of 2
larger for galaxies in \LR\,than for those in \HR. As for $\sigma_{\star,50}$, the bias 
decreases with increasing mass, but is present for many galaxies with
$M_{200}\lesssim 10^{12}\,{\rm M_\odot}$.\footnote{The quoted virial mass of $M_{200}\approx 10^{12}\,{\rm M_\odot}$ above
  which the sizes and velocity dispersions of galaxies in \LR\, appear unbiased relative to those in \HR\, seems
  conservative based on a visual impression of Fig.~\ref{fig:dscatter}. The bias in all three relations, though
  noticeable, is indeed quite small even at $M_{200}\approx 10^{11.5}\,{\rm M_\odot}$. In Section~\ref{ssec:conv} we
  provide a quantitative estimate of the halo mass above which spurious collisional effects at $r_{50}$ are
  unimportant, which for our \LR\, run is $M_{200}\approx 10^{11.7}\,{\rm M_\odot}$.}

\begin{figure}
  \includegraphics[width=0.48\textwidth]{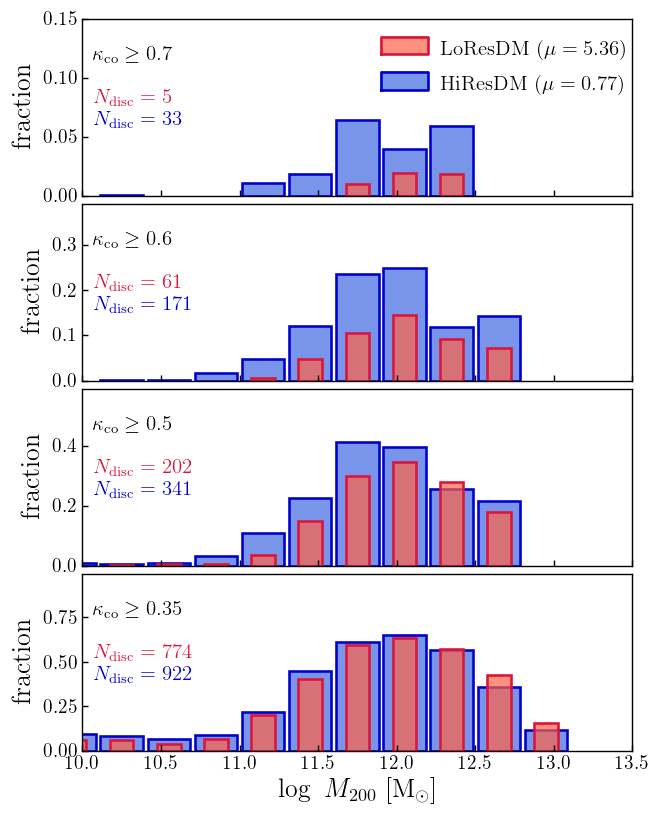}
  \caption{Fraction of central galaxies that are discs as a function of halo virial mass, $M_{200}$. From the bottom to top panel,
    discs are defined based on increasing $\kappa_{\rm co}$ thresholds, ranging from $\kappa_{\rm co}\geq 0.35$ (bottom
    panel) to $\kappa_{\rm co}\geq 0.7$ (top panel). The orange histograms show disc fractions in the \LR\,run and blue histograms
    show those in \HR. Note that the number of disc galaxies (which is quoted in each panel) is larger in
    \HR\,than in \LR, regardless of the $\kappa_{\rm co}$ threshold used.  For galaxies with $\kappa_{\rm co}\geq 0.5$,
    for which at least half of the stellar kinetic energy is in co-rotation, there are roughly 70 per cent more
    discs in \HR\,than in \LR.}
  \label{fig:fdisc}
\end{figure}

The spurious size growth, combined with the increased characteristic velocities of stellar particles, leads to an
artificial increase in the specific angular momentum of low-mass galaxies, as seen in the bottom panel of
Fig.~\ref{fig:dscatter}. This implies that, at sufficiently low mass, galaxies tend to {\em gain} angular momentum
from their surrounding DM haloes, which counters the findings of
\citet[][see also \citealt{Governato2004}]{Wilkinson2023}, who showed that thin galactic discs in idealised simulations
tend to {\em lose} angular momentum to the halo as a result of spurious heating. The angular momentum of
galaxies in cosmological simulations therefore exhibits a more complex dependence on spurious heating than what is captured by simple
idealised simulations of secular discs. The difference between our results and those of \citet{Wilkinson2023}
is partly because our cosmological
runs have a high prevalence of spheroidal galaxies (their results were based on the response of isolated discs to spurious
heating, which is different from that of dispersion supported spheroids), and partly due to the much larger spurious size growth
of galaxies in our cosmological runs.

These results have implications for simulated galaxy scaling relations, which we investigate below. 
Because these relations also depend on galaxy morphology, we will consider them separately
for disc galaxies (i.e. $\kappa_{\rm co}\geq 0.35$), ellipticals (i.e. $\kappa_{\rm co}\leq 0.25$),
and for the entire galaxy population. First, however,
we consider differences in the abundance of discs brought about by spurious heating.

\subsection{The fraction of galactic discs}
\label{ssec:fdisc}

\citet{Wilkinson2023} showed that spurious heating of simulated galactic discs affects not only the kinematics
of their stellar particles, but also their morphologies: discs become thicker as ordered
rotational motions are converted into random motions. In Fig.~\ref{fig:fdisc} we plot the fraction of
central galaxies that are discs as a function of their host halo virial mass, $M_{200}$.
Discs are defined as galaxies whose $\kappa_{\rm co}$ values 
exceed a series of thresholds that increase from the bottom to the top panel.
Note that in both simulations the total number of discs decreases with increasing $\kappa_{\rm co}$, but the {\em fraction} of discs
identified in \HR\, increases relative to that in \LR: e.g., for $\kappa_{\rm co}\geq 0.5$,
there are $\approx 70$ per cent more discs in \HR\,than in \LR; for $\kappa_{\rm co}\geq 0.7$, there are almost a factor of 7
more.

The impact of spurious heating on the morphologies of galaxies in cosmological simulations deserves a careful
analysis, which we defer to future work. In the sections that follow, however, we compare several standard galaxy
scaling relations obtained separately for discs (i.e. $\kappa_{\rm co}\geq 0.35$), ellipticals (i.e.
$\kappa_{\rm co}\leq 0.25$), and for the entire galaxy populations. 

\subsection{Implications for galaxy scaling relations}
\label{ssec:scaling}

\begin{figure*}
  \includegraphics[width=0.9\textwidth]{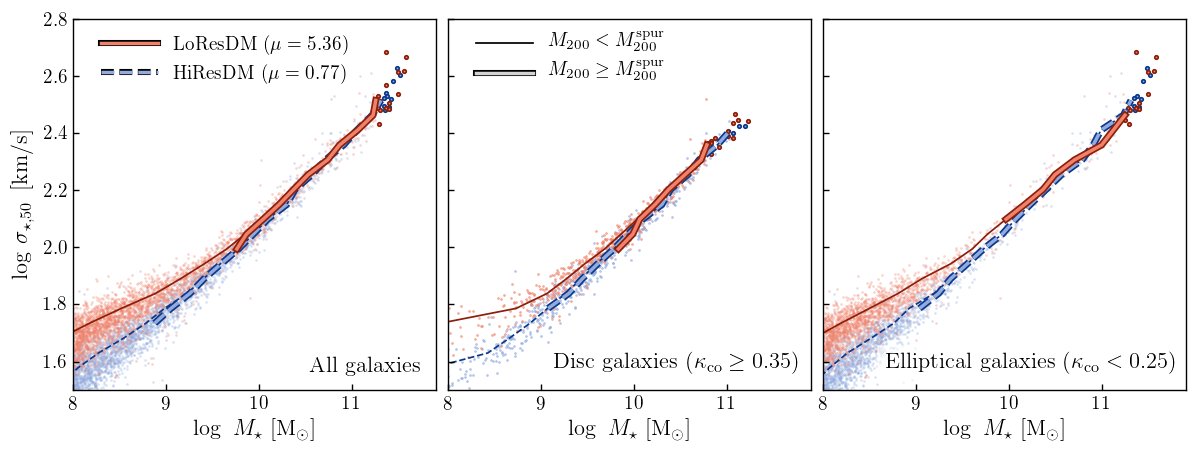}
  \caption{Total stellar velocity dispersion, $\sigma_{\star,50}$ (measured within a annular aperture
    of central radius $R=r_{50}$), of central galaxies plotted as a function of their stellar mass, $M_\star$.
    From left to right, different panels show results for the entire galaxy sample, and separately for
    disc ($\kappa_{\rm co}\geq 0.35$) and elliptical galaxies ($\kappa_{\rm co}\leq 0.25$), respectively.
    Thin coloured lines show the median relations obtained for all galaxies plotted in each panel (the
    individuals are shown as coloured dots); the thick lines show the median relations for the subsets of
    galaxies whose {\em measured} velocity dispersion, on average, exceeds the {\em spurious} velocity
    dispersion predicted for their halo mass using the empirical model of
    \citet[][see Section~\ref{ssec:conv} for details]{Ludlow2021}. As discussed by \citet{Ludlow2021},
    stellar particles in low-mass galaxies are vulnerable to spurious heating, which artificially increases
    their velocity dispersion.}
  \label{fig:sigMstar}
\end{figure*}

\subsubsection{The $\sigma_{\star,50} - M_\star$ relation}
\label{sssec:faber}

In Fig.~\ref{fig:sigMstar} we plot the characteristic velocity dispersion of central
galaxies $\sigma_{\star,50}$ as a function of their stellar mass, which serves as a proxy for the
Faber-Jackson (\citeyear{Faber1976}) relation.\footnote{Traditionally, the Faber-Jackson (\citeyear{Faber1976}) relation
is a projection of the fundamental plane of elliptical galaxies relating their central stellar velocity dispersion to
their total luminosity. In our analysis, $\sigma_{\star,50}$ includes contributions from both the random and ordered
motions of stellar particles in galaxies.}
Orange and blue lines show the median relations for galaxies in \LR\, and \HR, respectively; dots of
the corresponding colour show individual galaxies in the two runs. 
From left to right, different panels show results for the entire galaxy sample, and for the subsets of
discs and ellipticals, respectively.
As expected, low-mass galaxies ($M_\star \lesssim 10^{10}\,{\rm M_\odot}$) in \LR\, are kinematically
hotter than the ones in \HR, and the discrepancy increases with decreasing stellar mass, 
regardless of the galactic morphology. Above a characteristic mass of $M_\star\sim 10^{10}{\rm M_\odot}$, however,
the bias disappears and the two simulations are in good agreement.

The thick lines plotted in Fig.~\ref{fig:sigMstar} show the median relations for the subsets of galaxies whose
measured velocity dispersions,
$\sigma_{\star,50}$, exceeds the {\em spurious} velocity dispersion at $r_{50}$, i.e. $\sigma_{\rm spur}$, predicted for
their halo mass using the analytic model of
\citet[][with updates from \citealt{Wilkinson2023}]{Ludlow2021}, and were obtained as described in Section~\ref{ssec:conv}.
The median relations for these galaxies are in excellent agreement. 

\begin{figure*}
  \includegraphics[width=0.9\textwidth]{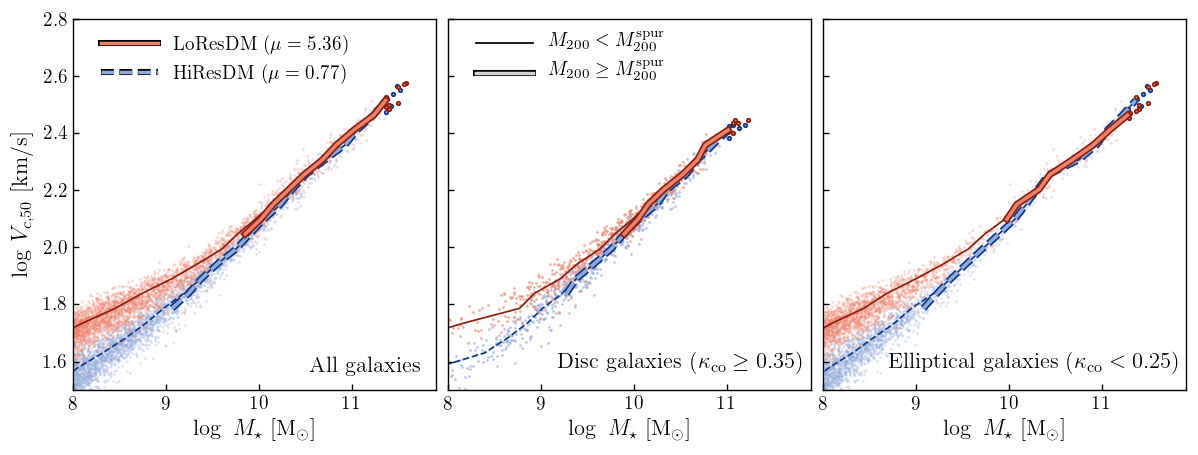}
  \caption{Same as Fig.~\ref{fig:sigMstar}, but for the rotational velocities of central galaxies (defined as the
    total circular velocity at the half-mass radius, $r_{50}$, i.e. $V_{c,50}\equiv V_c(r_{50})$). Colours,
    points, line styles and thicknesses have the same meaning as in Fig.~\ref{fig:sigMstar}. }
  \label{fig:vrotMstar}
\end{figure*}

\subsubsection{The $V_{\rm c,50} - M_\star$ relation}
\label{sssec:tully}

As for the stellar velocity dispersion, the rotational velocities of galaxies -- approximated here as the 
total circular velocity at $r_{50}$, i.e. $V_{c,50}$ -- are also affected by the mass resolution of the DM component.
Fig.~\ref{fig:vrotMstar} plots the relation between $V_{c,50}$ and $M_\star$, a proxy for the Tully-Fisher~(\citeyear{TF1977})
relation, for our entire sample of central galaxies (left panel), as well as for late- (middle panel) and early-types (right panel),
using the same plotting conventions as in Fig.~\ref{fig:sigMstar}. 

Like their velocity dispersions, the rotation velocities of low-mass galaxies differ systematically between
the runs: those in \LR\, typically have higher $V_{c,50}$ values than those in \HR, but
the discrepancy disappears for massive, well-resolved galaxies.
Indeed, thick lines in Fig.~\ref{fig:vrotMstar} show the median relations for the same subsets of galaxies used for
Fig.~\ref{fig:sigMstar}, i.e. the those whose velocity dispersion exceeds the {\em spurious} velocity dispersion
predicted for their halo mass (see Section~\ref{ssec:conv}). Regardless of morphological type, galaxies confined to these
samples have unbiased rotational velocities.
The systematic difference in the $V_{c,50}$ values of low-mass galaxies is 
partly due to their larger characteristic radii, $r_{50}$ (see Fig.~\ref{fig:dscatter}),
which probe the rising part of the $V_c(r)$ profiles, but also due to systematic differences in their inner
stellar and DM mass profiles. We briefly discuss the latter point in Section~\ref{ssec:res}.

\begin{figure*}
  \includegraphics[width=0.9\textwidth]{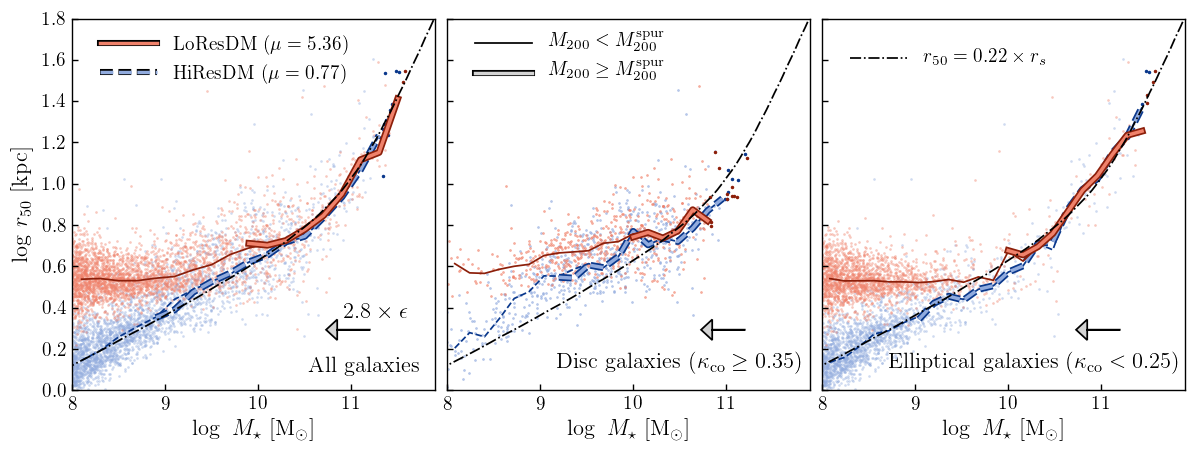}
  \caption{Size-mass relations for all central galaxies (left panel), and for the subset of disc
    ($\kappa_{\rm co}\geq 0.35$; middle panel) and elliptical galaxies ($\kappa_{\rm co}\leq 0.25$;
    right panel). Sizes are quantified by the three-dimensional half-stellar mass radius, $r_{50}$.
    As discussed by \citet{Ludlow2019}, low-mass galaxies are vulnerable to
    collisional heating and experience spurious size growth. Good convergence is obtained for systems in which
    spurious heating is sub-dominant (the thick lines; see Section~\ref{ssec:conv} for details). The dot-dashed
    black lines in each panel show the
    empirical relation $r_{50}=0.22\times r_{\rm s}$, where $r_{\rm s}$ is the scale radius of an NFW halo of virial
    mass $M_{200}$ that hosts a galaxy of mass $M_\star$ inferred from the best-fitting ${\rm M_\star - M_{200}}$
    relation plotted in Fig.~\ref{fig:ms_mbh_mh} (we use the $r_{\rm s}-M_{200}$ relation of \citealt{Ludlow2016}).
    The spline softening length, i.e. 
    $\epsilon_{\rm sp}\equiv 2.8\times \epsilon=1.96\,{\rm kpc}$, is show using a grey arrow in each panel.}
  \label{fig:sizes}
\end{figure*}

\begin{figure}
  \includegraphics[width=0.48\textwidth]{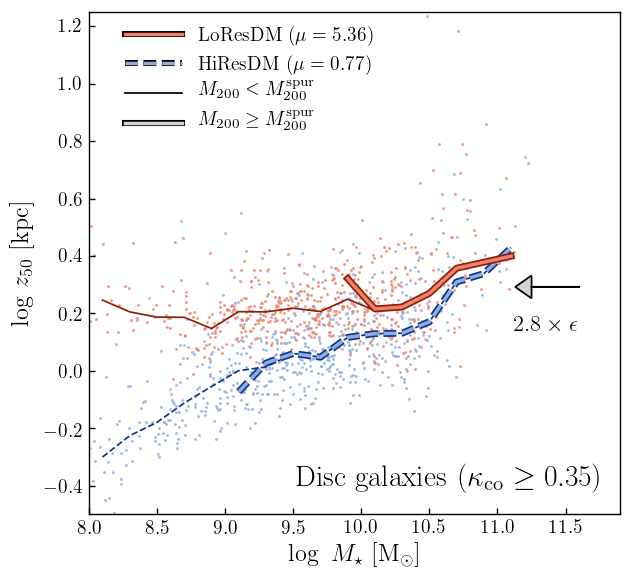}
  \caption{Vertical half-stellar mass height, $z_{50}$ (measured within a cylindrical annulus of central radius $R=r_{50}$
    and width $\Delta\log R=0.3$), for disc galaxies ($\kappa_{\rm co}\geq 0.35$) plotted as a function of stellar mass.
    Colours and line styles/weights have the same meaning as in Fig.~\ref{fig:sizes}.}
  \label{fig:z50}
\end{figure}

\subsubsection{The $r_{50} - M_\star$ relation}
\label{sssec:size}

The size-mass relations obtained for central galaxies are plotted in Fig.~\ref{fig:sizes}.
Regardless of morphological type, the two relations agree at high stellar masses
(i.e. $M_\star \gtrsim 10^{10}{\rm M_\odot}$), but toward low
masses the median curves veer away from one another, where the median sizes of galaxies in \LR\,
approach a constant value of $r_{50}\approx 3.5\,{\rm kpc}$ while those in \HR\, continue to decline.
These results confirm and extend those previously obtained by \citet{Ludlow2019}.

It is tempting to associate the plateauing sizes of low-mass galaxies in \LR\, with the gravitational
softening length ($\epsilon_{\rm sp}$ is indicated by arrows in Fig.~\ref{fig:sizes}). However,
sizes in \HR\, -- which employed the {\em same} softening length as \LR\,--
decline monotonically with decreasing mass, reaching values that are considerably smaller than 
$\epsilon_{\rm sp}=2.8\times\epsilon$.
The gravitational softening length therefore does not artificially increase the sizes of low-mass
galaxies in our runs,\footnote{\citet{Ludlow2020} showed that the gravitational softening length $\epsilon$ does, in
fact, affect sizes of galaxies; at fixed $m_{\rm DM}$ and $m_{\rm gas}$, small values of $\epsilon$ exacerbate 2-body
scattering, leading to systematically larger galaxy sizes, while large values of $\epsilon$ suppress the effect.} 
suggesting that other effects are at work. Our explanation, already emphasised by
\citet[][see also \citealt{Revaz2018}]{Ludlow2019}, is that spurious heating results in
artificial size growth. 

Support for our interpretation is provided by the thick lines in Fig.~\ref{fig:sizes}, which show
the median size-mass relations obtained for the subset of galaxies whose velocity dispersions, $\sigma_{\star,50}$,
exceed the spurious velocity dispersion at $r_{50}$ expected for their halo mass (see Section~\ref{ssec:conv}).
In addition to their characteristic velocities,
$\sigma_{\star,50}$ and $V_{c,50}$, our empirical model offers a simple way to identify galaxies whose characteristic
sizes are absolved of spurious collisional effects.

Finally, note that the median size-mass relation in \HR\, (and the relation for well-resolved galaxies
in \LR) are accurately approximated by the simple empirical relation\footnote{This is a slightly modified version of the
  empirical relation between the half-mass radii of galaxies and the scale radii of their DM haloes proposed
  by \citet{Navarro2017}, namely $r_{50}=0.2\times r_{\rm s}$. Although the origin of this relation in \eagle\,
  is unclear, it proves useful when modelling the spurious heating rates for our simulated galaxy populations,
  as described in Section~\ref{ssec:conv}. We acknowledge, however, that the size-mass relation of discs differs from 
  that of spheroids in both simulations \citep[e.g.][]{Rodriguez-Gomez2022} and observations \citep{Huang2017}. }
\begin{equation}
  r_{50}=0.22\times r_{\rm s},
  \label{eq:r50rs}
\end{equation}
where $r_{\rm s}$ is the NFW scale radius of the DM halo hosting the galaxy (plotted using
standard scaling laws between $M_{200}$ and concentration, see \citealt[][]{Ludlow2016}, and using our best-fitting
$M_\star - M_{\rm halo}$ relation).
Equation~(\ref{eq:r50rs}) is shown as a dot-dashed line in each panel of Fig.~\ref{fig:sizes}.

\subsubsection{The $z_{50} - {\rm M_\star}$ relation}
\label{sssec:heights}

The stellar half-mass heights, $z_{50}$, for the subset of disc galaxies, are plotted 
in Fig.~\ref{fig:z50} (using the same conventions as Fig.~\ref{fig:sizes}).
As for the other structural and kinematic properties of galaxies, the scale heights of massive galaxies
($M_\star\gtrsim 10^{10}{\rm M_\odot}$)
in our two simulations are in best agreement, whereas towards low masses the agreement worsens:
for $M_\star\lesssim 10^{10}{\rm M_\odot}$,
the half-mass heights in \LR\, are approximately constant ($z_{50}\approx \epsilon_{\rm sp} \approx 2\, {\rm kpc}$),
whereas galaxies in \HR\, become progressively thinner relative to the softening length.
As for $r_{50}$, good (though not perfect) agreement in the scale heights of discs is obtained for those deemed
unaffected by spurious heating. The median relations for these galaxies are highlighted in Fig.~\ref{fig:z50}
using thick lines.

\subsubsection{The $j_\star - M_\star$ relation}
\label{sssec:Fall}

\begin{figure*}
  \includegraphics[width=0.9\textwidth]{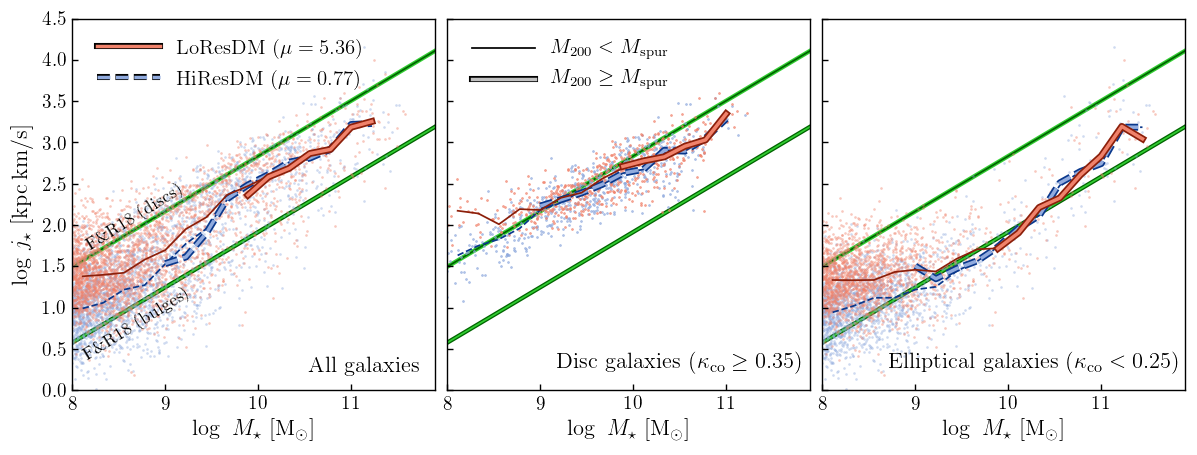}
  \caption{Total specific angular momentum as a function of stellar mass
    (both quantities were calculated using all stellar particles bound to each central galaxy). The different 
    panels show results for the full galaxy population (left), and for the subset of disc (middle;
    $\kappa_{\rm co}\geq 0.35$) and elliptical galaxies (right; $\kappa_{\rm co}\leq 0.25$).
    The solid green lines in each panel show the relations obtained by \citet{FallRoman2018} for observed
    discs and bulges. Other plotting conventions follow from Fig.~\ref{fig:sigMstar}.}
  \label{fig:fall}
\end{figure*}

The final kinematic scaling relation we consider is the \citet{Fall1983} relation between specific angular momentum 
and stellar mass, which we plot in Fig.~\ref{fig:fall}.
Plotting conventions (e.g. line styles, weights, and colours) and layout were inherited 
from previous figures.

As shown in Fig.~\ref{fig:dscatter}, the specific angular momentum of massive galaxies in our runs
are in good agreement, whereas at low masses the angular momentum of galaxies in \LR\, typically exceeds that
of galaxies in \HR. Although idealised discs lose angular momentum to their surrounding DM
haloes as a result of gravitational scattering \citep[e.g.][]{Governato2004,Wilkinson2023},
their stellar particles tend to gain energy by the same process, which propels them onto higher
energy orbits, increasing the characteristic velocity and size of the galaxy. The latter effect also occurs in
dispersion supported systems, which dominate the low-mass galaxy population in our cosmological runs, 
increasing their angular momentum. As a result, low-mass galaxies in \LR\,have systematically {\em more} angular
momentum than those in \HR\, with the same stellar mass.

These compensating effects reduce the galaxy mass scale above which the $j_\star - M_\star$
relations in our runs agree with one another. The thin and thick lines in Fig.~\ref{fig:fall}
have the same meaning as in the previous two plots and show, respectively, the median relations for the entire samples of
galaxies and for the subsets whose velocity dispersions, $\sigma_{\star,50}$, are predicted to be free from spurious
heating. Note that the $j_\star - M_\star$ relations for the subsets of disc and elliptical galaxies 
do not depend on the DM mass resolution for stellar masses $M_\star \gtrsim 10^9 {\rm M_\odot}$ (corresponding to halo masses
$M_{200}\gtrsim 10^{11.1}{\rm M_\odot}$).

Finally, note that the difference in the angular momentum of low-mass galaxies that arises due to
spurious heating is much smaller than the intrinsic difference in the angular momentum content of discs and spheroids,
as inferred from both simulations and observations.
The observed relations are shown in Fig.~\ref{fig:fall} as green lines, and correspond to the best-fitting
relations for discs and bulges obtained by \citet{FallRoman2018}. These relations differ at all masses by about an order
of magnitude, whereas the median angular momentum of $M_\star\lesssim 10^9{\rm M_\odot}$ galaxies in our two
simulations differ by much less  (with even better agreement at higher stellar masses). 

\begin{figure*}
  \includegraphics[width=0.9\textwidth]{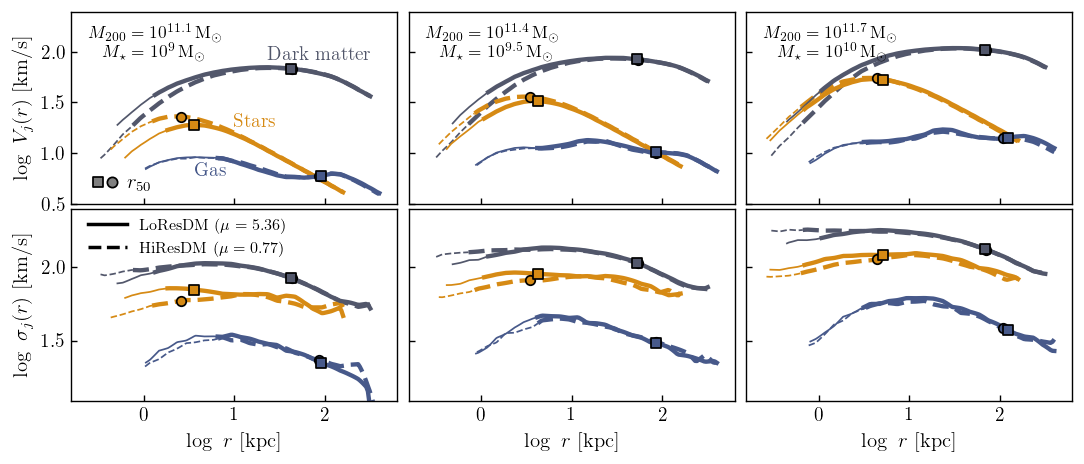}
  \caption{Top panels plot the contribution to the total circular velocity profile due to dark matter (dark grey curves),
    stellar (yellow curves) and gas particles (blue curves); bottom panels plot velocity dispersion profiles of each component
    (for gas particles, we plot the one-dimensional velocity dispersion perpendicular to the plane of the gaseous disc, which is
    the component most susceptible to spurious heating).
    From left-to-right, the three columns correspond to galaxies (and their DM haloes) with different stellar masses:
    $M_\star=10^9\,{\rm M_\odot}$, $10^{9.5}\,{\rm M_\odot}$, and $10^{10}\,{\rm M_\odot}$, respectively. The corresponding average
    halo masses, $M_{200}$, are also indicated, and do not differ between the two runs. In all panels, results from \LR~and \HR~are
    plotted using solid and dashed lines, respectively, with thin (thick) lines extending to the smallest radii enclosing 10 (100)
    particles of each species. Circles or squares marked along each curve indicate the location of the half-mass radius
    of each component. The values of stellar mass chosen for this plot are comparable to the lower mass limits imposed by various
    authors on the analysis of simulations with mass and force resolution comparable to ours (see text for details).}
  \label{fig:Vc}
\end{figure*}

\section{Implications of spurious heating for cosmological simulations of galaxy formation}
\label{sec:imp}

\subsection{General considerations}
\label{ssec:gen}

In simulations of galaxy formation gas particles experience radiative energy losses and condense in
the inner regions of DM haloes. Under the right conditions, stellar particles form from these gas
particles, inheriting their kinematics and their masses. The baryonic components of galaxies
are thus initially colder than their surrounding DM haloes. 

Gravitational scattering leads to energy exchanges between DM and baryonic particles and a tendency
toward energy equipartition. 
When $\mu>1$, the lighter stellar particles heat up and the heavier DM particles cool down as energy
exchanges strive to equalise their mean kinetic energies. This form of heating occurs even when the 
lighter and heavier particles are initially well mixed, and it leads to mass segregation. 

Another form of heating occurs because stellar particles are initially colder than the DM. This effect is
different from mass segregation: it occurs for {\em all} $\mu\geq 1$. For the particular case $\mu=1$,
the rate of heating by this mechanism depends on the relaxation time and hence the total number of particles
by which a galaxy is resolved (i.e. at fixed mass, it depends on $m_{\rm DM}$). But the maximum
amount of heating is dictated by the amount of dissipation the baryons experienced. If the dissipation is
large, the spurious heating can also be large, but if the dissipation is negligible, then the spurious
heating will also be negligible.

The common situation in galaxy formation simulations has both energy dissipation in the baryons and
differences in stellar and DM particle masses ($\mu\approx  5$), thus leading to a more complicated situation
in which the rate of spurious heating depends on both particle masses (i.e. on $m_{\rm DM}$ and $\mu$),
and on the amount of gaseous dissipation (i.e. on the initial separation of DM and
stellar particles in phasespace). Although both effects heat stellar particles and cool DM particles, in practice we find
that spurious heating is dominated by the the amount of dissipation, even when $\mu>1$. This explains the weak $\mu$-dependence of disc
heating rates reported by \citet{Ludlow2021}.

We must acknowledge that the internal structure and kinematics of all galaxies in cosmological simulations are 
affected by spurious heating, and it is likely that the properties of their DM haloes are as well, at least
in their centres. In the sections that follow, we show that this is indeed the case and offer some guidance on how
to disentangle spurious results from robust ones. 

\subsection{Resolving the inner structure of galaxies and their dark matter haloes}
\label{ssec:res}

Assessing numerical convergence in hydrodynamical simulations is complex, particularly when the spatial scales relevant for
feedback from stars and AGN are not resolved, as is usually the case. Defining samples of ``well-resolved'' galaxies 
is therefore difficult and rarely done convincingly. For example, past studies using
\eagle\, have adopted different lower limits on galaxy stellar masses ($M_{\star,{\rm lim}}$, above which results are {\em assumed} to be
robust), even when carrying out similar analyses.
Some have adopted $M_{\star,{\rm lim}}\approx 10^9\,{\rm M_\odot}$ \citep[e.g.][corresponding to several
  hundred stellar particles]{Furlong2017,Ferrero2017}, while others used $M_{\star,{\rm lim}}\approx 10^{10}\,{\rm M_\odot}$
\citep[corresponding to several thousand stellar particles, e.g.][]{Correa2017,Ferrero2021,deGraaff2022}; 
others chose values in between \citep[e.g.][]{Lagos2017,vandeSande2019,Thob2019}.
For IllustrisTNG-100, which has a mass and spatial resolution comparable to that of \eagle, the situation is similar.

Motivated by this, in Fig.~\ref{fig:Vc} we plot radial profiles of the average circular velocity
(upper panels) and velocity dispersion (lower panels) for galaxies with 
$M_\star \approx 10^9\,{\rm M_\odot}$, ${\rm 10^{9.5}\,M_\odot}$, and ${\rm 10^{10}\,M_\odot}$, which roughly span the
range of $M_{\star,{\rm lim}}$ values mentioned above (left to right panels, respectively; note that we use a bin width
of $\Delta\log M_\star = 0.3$). 
The profiles are plotted separately for DM (grey curves), stellar (yellow curves) and gas particles (blue curves;
for gas particles we plot the velocity dispersion perpendicular to the disc plane),
using solid lines for \LR\, and dashed lines for \HR. The thick and thin lines
are plotted to minimum radii that enclose, on average, 100 and 10 particles of each type, respectively;
the half-mass radii are indicated by symbols along each curve (squares for \LR; circles for \HR). These profiles clarify the
results reported in previous sections, and highlight several others.

First, the radial profiles for gas closely overlap at all radii, even when only  $\approx 10$
particles are enclosed. Artificial heating of gas particles by DM particles is therefore unimportant
in our simulations (but likely is important for simulations that use very massive DM particles; e.g.
\citealt{Steinmetz1997}). For gas particles, radiative cooling dissipates the energy gained by spurious heating,
which explains why the {\em global} properties of galaxies in our runs (e.g.
the total baryonic masses, SFHs, etc), are in good agreement, even when the internal structure and
kinematics of their stellar components are not.

There is also good agreement between the {\em outer} $V_j(r)$ and $\sigma_j(r)$ profiles
for stellar and DM particles. However, systematic differences are evident at small radii
that become more apparent toward low masses. Specifically, galaxies in \LR\, have more extended and kinematically hotter
stellar components, but more concentrated and (slightly) cooler central DM distributions.\footnote{Differences in the concentrations
of the stellar or DM distributions shown in Fig.~\ref{fig:Vc} can be inferred by comparing their circular velocities at fixed
radii: higher/lower $V_j(r)$ implies a higher/lower concentration.} Note that differences in the $V_\star(r)$ and
$\sigma_\star(r)$ profiles at fixed mass extend to radii that exceed $r_{50}$, suggesting that differences
in the $\sigma_{\star,50}-M_\star$ and $V_{c,50}-M_\star$ relations obtained from our simulations (Figs.~\ref{fig:sigMstar}
and~\ref{fig:vrotMstar}, respectively)
are not merely due to differences in the radius $r_{50}$ at which these characteristic velocities were measured. 

Note too that the results plotted in Fig.~\ref{fig:Vc} differ from the expectations of simple adiabatic contraction models,
which predict that more compact galaxies give rise to higher central DM densities. Our results, in fact, show the opposite trend,
and one that is consistent with the expectations of spurious heating. As discussed by
\citet[][see also \citealt{Ludlow2020}]{Ludlow2019}, energy is more readily transferred
from DM to stellar particles by 2-body scattering when $\mu \gtrsim 1$, and the net effect is for galaxies to heat-up and expand,
while DM haloes cool down and contract (i.e. $\sigma_{\rm DM}$ decreases at fixed $r$ and $V_{\rm DM}$ increases at fixed $r$).
This effect is apparent in all panels of Figure~\ref{fig:Vc}. Unless great care is taken to eliminate galaxies affected by spurious heating,
cosmological simulations should therefore not be used to develop or calibrate empirical models for the baryon-induced
contraction of DM haloes.

Finally, note that for all mass bins plotted in Figure~\ref{fig:Vc}, the circular velocity profiles of DM haloes are
{\em not converged} at radii $r\lesssim r_{50}$ (i.e. below the stellar half-mass radius),
even when their stellar and gaseous $V_c(r)$ profiles are converged.
This has important implications for modelling the rotation curves and DM fractions of the least-massive galaxies in
hydrodynamical simulations, which are also sensitive to spurious heating. Models for predicting the ``convergence
radius'' of DM haloes, such as those advocated by \citet{Power2003} and \citet{Ludlow2019b}, will require revision
to accommodate these results. 

\hspace{1cm}
\subsection{A convergence criterion for spurious heating in cosmological hydrodynamical simulations}
\label{ssec:conv}

\setlength{\tabcolsep}{0.45em}

\begin{center}
  \begin{table}
    \caption{Values of $M_{200}^{\rm spur}$ at $z=0$ required for convergence in galactic structure and kinematics
      at the half stellar mass radius, $r_{50}$, for various cosmological simulations. The values for \HR\, were
      estimated using the median $\sigma_{\star,50} - M_{200}$ relation in that run (see Fig.~\ref{fig:sigm200}), and
      are provided for several values of the DM particle mass. The value obtained for the 100 (25) cubic Mpc \eagle \,
      run is provided in the row labelled \eagle-100 (\eagle-25); rows labelled TNG-100 and TNG-50 provide the
      $M_{200}^{\rm spur}$ values for the IllustrisTNG-100 and IllustrisTNG-50 simulations, respectively.}
    \begin{tabular}{r r c c c c r r}\hline \hline
      & Sim. & $N_{\rm DM}/N_{\rm gas}$ & $m_{\rm DM}$             & $\mu$ & $M_{200}^{\rm spur}$       & $N_{200}^{\rm spur}$ &\\
      &   &     & $[10^5\,{\rm M_\odot}]$  & $m_{\rm DM}/m_{\rm gas}$ & $[10^{10}\,{\rm M_\odot}]$ & $[10^4]$             &\\\hline
      & \HR & 1     &     97.0           &   5.36   & 54.9          &   5.7          & \\
      & \HR & 2     &     48.5           &   2.68   & 36.2          &   7.5          &\\
      & \HR & 3     &     32.3           &   1.79   & 28.1          &   8.7          &\\
      & \HR & 4     &     24.2           &   1.34   & 23.7          &   10.0         &\\
      & \HR & 5     &     19.4           &   1.07   & 20.7          &   10.6         &\\
      & \HR & 6     &     16.2           &   0.89   & 18.5          &   11.4         &\\
      & \HR & 7     &     13.9           &   0.77   & 16.8          &   12.1         &\\\space
      & \eagle-100  & 1 & 97.0           &   5.36   & 54.9          &   5.1         & \\
      & \eagle-25   & 1  &12.2           &   5.36   & 9.91          &   8.2         & \\ 
      & TNG-100     & 1  &75.0           &   5.36   & 50.1          &   9.9         & \\ 
      & TNG-50      & 1  &4.50           &   5.36   & 7.9  &  26.1         & \\ \hline

    \end{tabular}
    \label{TabMspur}
  \end{table}
\end{center}

\citet[][and later \citealt{Wilkinson2023}]{Ludlow2021} derived and tested a useful approximation for the evolution of
$\sigma_i$ (the three cylindrical components of the stellar velocity dispersion, i.e. $\sigma_{\rm R}$, $\sigma_\phi$,
and $\sigma_z$) that arises due to the spurious heating of simulated galactic discs by DM halo particles.
This model is
\begin{equation}
  \sigma^2_i=\sigma_{\rm DM}^2 - (\sigma_{\rm DM}^2-\sigma_{i,0}^2)\times \exp\biggr(-\frac{t}{t_{\sigma_i}}\biggl),
  \label{eq:exp}
\end{equation}
where
\begin{equation}
  t_{\sigma_i}=\frac{V_{200}^3}{G^2\rho_{\rm DM} m_{\rm DM}}\biggr[\sqrt{2}\,\pi\,k_i\, \ln\Lambda \biggr(\frac{\rho_{\rm DM}}{\rho_{200}}\biggl)^{\alpha_i}\biggr(\frac{V_{200}}{\sigma_{\rm DM}}\biggl)^2\, \biggl]^{-1}
  \label{eq:tvir}
\end{equation}
is a characteristic ``heating'' timescale at which $\sigma_i\sim\sigma_{\rm DM}$ (the local 1-dimensional
velocity dispersion of DM), and $\sigma_{i,0}$ is the initial velocity dispersion in the $i$ direction.\footnote{Equation~(\ref{eq:exp})
above, which is equivalent to equation~(6) in \citet{Wilkinson2023}, 
is motivated by the disc heating rates derived analytically by \citet{LO1985}. In their case, the parameters
$\alpha_i=0$ and $k_i \ln\Lambda$ in equation~(\ref{eq:tvir}) can be calculated explicitly
when a velocity distribution for DM particles is specified. In this paper, we adopt the parameter values listed
in Table 2 of \citet{Wilkinson2023}, i.e. $(\alpha_{\rm R},\alpha_z,\alpha_\phi)=(-0.189,-0.308,-0.115)$ and
$(k_{\rm R},k_z,k_\phi)\ln\Lambda =(20.17,20.19,9.40)$. We refer the interested reader to \citet{Ludlow2021} and
\citet{Wilkinson2023} for a detailed discussion of our disc heating model.}

For $\sigma_{i,0}\ll \sigma_i\ll\sigma_{\rm DM}$, equation~(\ref{eq:exp}) reproduces the
disc heating rates derived analytically by \citet{LO1985}. But for $t\gg t_{\sigma_i}$, $\sigma_i$ approaches
$\sigma_{\rm DM}$, which reproduces the results of idealised simulations of poorly-resolved galactic discs on
timescales of order or less than the ages of galaxies of interest ($t\lesssim 10\,{\rm Gyr}$).

The total spurious velocity dispersion predicted by equation~(\ref{eq:exp}) is given by
\begin{equation}
  \sigma^2_{\rm spur}=\sum_i \sigma^2_i=\sigma_{\rm R}^2 + \sigma_\phi^2 + \sigma_z^2,
\end{equation}
and depends on halo mass through $V_{200}$, and on galacto-centric radius through
$\sigma_{\rm DM}$ and $\rho_{\rm DM}$. The latter, $\rho_{\rm DM}$, is fully specified by the cosmological parameters 
\citep[see e.g.][]{Ludlow2016,Lopez-Cano2022} and $\sigma_{\rm DM}$ can be obtained from Jeans'
equations, allowing predictions to be made for the level of spurious heating a simulated galaxy
may have undergone given its age and halo mass.\footnote{A python script to calculate the spurious disc
heating rate based on equation~(\ref{eq:exp}) can be found at~\url{https://github.com/AaronDLudlow/hot-disc.git}.}

\begin{figure}
  \includegraphics[width=0.48\textwidth]{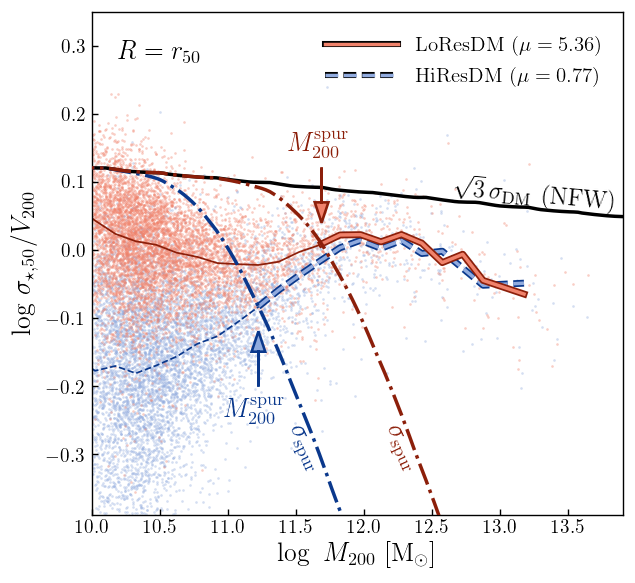}
  \caption{Stellar velocity dispersion $\sigma_{\star,50}$ at $r_{50}$ for
    central galaxies plotted as a function $M_{200}$ (and normalised by $V_{200}$). Orange and blue colours distinguish
    our \LR\,and \HR\,runs, respectively. The solid black line shows the total DM velocity dispersion
    at $r_{50}$ for an isotropic NFW halo (assuming $r_{50}=0.22\times r_{\rm s}$ where $r_{\rm s}$ is
    the halo scale radius; see Fig.~\ref{fig:sizes}).
    The two dot-dashed lines labelled $\sigma_{\rm spur}$ show the spurious velocity dispersion
    predicted for our runs using equation~(\ref{eq:exp}), assuming a galaxy age of $t_\star = 7.6\,\,{\rm Gyr}$ (approximately
    the median half-mass stellar age of galaxies with ${\rm M_\star\geq 10^9\,M_\odot}$). Individual galaxies
    are shown as small dots and the median relations as coloured lines. The
    arrows labelled $M_{200}^{\rm spur}$ indicate the halo masses at which $\sigma_{\rm spur}$
    intersects the median velocity dispersion-stellar mass relation of galaxies in each simulation. These median relations
    are shown using thick and thin lines, respectively, for $M_{200}\geq M_{200}^{\rm spur}$
    and $M_{200}\leq M_{200}^{\rm spur}$.}
  \label{fig:sigm200}
\end{figure}

In Fig.~\ref{fig:sigm200}, we compare the predictions of equation~(\ref{eq:exp})
for the total spurious stellar velocity dispersion (dot-dashed lines) to the values of $\sigma_{\star,50}$,
measured in our simulations. Both dispersions are normalised by $V_{200}$ and are plotted as a function
of $M_{200}$. Note that our model assumes $\sigma_{i,0}=0$ and a stellar half-mass radius of $r_{50}=0.22\times r_{\rm s}$
(see Fig.~\ref{fig:sizes}; $r_{\rm s}$ was obtained from the mass-concentration
relation of \citealt{Ludlow2016}), and we adopt a typical age $t=7.6\,{\rm Gyr}$, roughly the median half-mass
age of \eagle\, galaxies with $M_\star\geq 10^9{\rm M_\odot}$.

The $\sigma_{\star,50}-M_{200}$ relations from our simulations evoke trends seen in previous figures;
above a characteristic halo mass they coincide, but below that mass they differ.\footnote{Note that 
the lowest-mass galaxies plotted in Fig.~\ref{fig:sigm200} are not maximally heated, as anticipated by equation~(\ref{eq:exp}).
This may be because
1) their stars are not mono-age (younger stars are less affected by spurious heating than older ones); or 2) the
effects of spurious heating are suppressed by the gravitational force softening, which is only slightly smaller than
the half-mass radii of the lowest mass galaxies \citep[see, e.g.,][]{Ludlow2019}.} Importantly, this characteristic halo mass
roughly coincides with the mass above which the median velocity dispersion
of the galaxies in \LR\, first exceeds the {\em spurious} velocity dispersion $\sigma_{\rm spur}$ predicted
by equation~(\ref{eq:exp}).\footnote{We have verified that similar
  results are obtained when this comparison is limited to any of the three orthogonal components of the velocity
  dispersion, i.e. $\sigma_{\rm R}$, $\sigma_z$, and $\sigma_\phi$.} This halo mass 
is labelled $M_{200}^{\rm spur}$ in Fig.~\ref{fig:sigm200} and is pinpointed by coloured arrows (orange for \LR\, and blue for \HR). For
both simulations, the median $\sigma_{\star,50}- M_{200}$ relations 
are represented by thick line segments for $M_{200}\geq M_{200}^{\rm spur}$, where they are in good agreement.
At $M_{200}^{\rm spur}$, the median relations differ by $\approx 0.03$ dex. 

Other fundamental properties of galaxies occupying haloes with masses $M_{200}\geq M_{200}^{\rm spur}$ are also converged,
specifically those plotted in Figs.~\ref{fig:sigMstar} through~\ref{fig:z50}, where we also used thick and thin lines
to distinguish the median relations for galaxies with $M_{200}\geq M_{200}^{\rm spur}$ and
$M_{200}<  M_{200}^{\rm spur}$, respectively. 
Fig.~\ref{fig:sigm200} therefore summaries our main result: {\em the kinematic and structural 
properties of galaxies are robust to spurious heating provided the intrinsic velocity dispersion
  of their stellar particles exceeds the spurious velocity dispersion predicted by equation~(\ref{eq:exp}).}

Although the velocity dispersions plotted in Fig.~\ref{fig:sigm200} were measured at $r_{50}$,
we show in Appendix~\ref{appB} that equation~(\ref{eq:exp}) can also be used to assess whether spurious heating
is important at other galacto-centric radii and for mono-age stellar populations, for which different levels of
heating are expected. 

\section{Summary}
\label{sec:summary}

We have used two cosmological, smoothed-particle hydrodynamical simulations based on the \eagle\,model of galaxy
formation to study the impact of spurious heating of stellar particles by DM particles on the properties of
simulated galaxies. Our simulations differ only in the mass of their DM particles (those in \HR\, are a factor
of 7 less massive than those in \LR), but share all other numerical and subgrid parameters, including the baryonic
particle mass and force resolution. The main features of the simulations are described in Section~\ref{SecSims}
and Table~\ref{TabSimParam}. Our main results are summarised below.

\begin{enumerate}

\item In Section~\ref{sec:galpop} we identified several global properties of the galaxy population that
  are not affected by spurious collisional heating: their total stellar, gas, and black hole (BH)
  masses, and the virial masses of their dark matter (DM) haloes (Fig.~\ref{fig:mscatter}). When matched
  between the runs, the stellar and gas masses of individual galaxies do, however, exhibit non-negligible
  scatter that increases with decreasing halo mass. This is usually attributed to the stochastic star
  formation and feedback models often adopted for cosmological simulations~\citep[e.g.][]{Borrow2022b},
  but may also arise from the spurious numerical effects discussed in this paper. The scatter for BH
  masses, decreases with decreasing halo mass. This is because the majority of low-mass haloes
  (${\rm M_{200}\lesssim 10^{11} M_\odot}$) contain only one central BH particle, whose mass is a small
  multiple of the BH seed mass. The virial masses of individual haloes typically agree to better than 10
  per cent. The good agreement between the baryonic masses of galaxies stems from similarities in
  their star formation histories (Fig.~\ref{fig:sfh}).      

\item As a result, both simulations have similar stellar-halo and stellar-BH mass relations
  (Fig.~\ref{fig:ms_mbh_mh}), as well as similar stellar, BH, and DM halo mass functions (for both central
  and satellite galaxies; Fig.~\ref{fig:mf}). Once the subgrid parameter values have been chosen for a
  simulation such that it reproduces observations of the galaxy stellar mass function, or the observed
  relation between stellar and BH masses, they should therefore not require re-calibration if the DM mass
  resolution is increased such that $\mu\approx 1$.

\item In Section~\ref{sec:scaling} we showed that the structural and kinematic properties of the stellar
  components of simulated galaxies {\em are} affected by spurious collisional heating. For example, the
  fraction of disc galaxies (whose $\kappa_{\rm co}$ values exceed various thresholds) is higher in \HR\,
  than in \LR\, (Fig.~\ref{fig:fdisc}), in qualitative agreement with predictions of \citet{Wilkinson2023}. 
  Spurious heating also artificially increases the stellar velocity dispersion (Fig.~\ref{fig:sigMstar}),
  rotation velocity (Fig.~\ref{fig:vrotMstar}), size (Fig.~\ref{fig:sizes}), thickness (Fig.~\ref{fig:z50}),
  and angular momentum (Fig.~\ref{fig:fall}) of low-mass galaxies; in all cases, the bias between \LR\, and
  \HR\, increases with decreasing mass. The structural and kinematic properties of the gaseous components of
  galaxies are not affected by spurious collisional heating, at least for the DM particle masses we study in
  this paper. 

\item As a result, spurious collisional heating changes the normalisation and slope of simulated galaxy scaling
  relations, unless efforts are made to eliminate systems that have been adversely affected by it. The
  Faber-Jackson (\citeyear{Faber1976}) relation between velocity dispersion and stellar mass 
  (Fig.~\ref{fig:sigMstar}), the Tully-Fisher (\citeyear{TF1977}) relation between rotation velocity and stellar
  mass (Fig.~\ref{fig:vrotMstar}), and the \citet{Fall1983} relation between angular momentum and stellar mass
  (Fig.~\ref{fig:fall}) are all sensitive to spurious heating, as are structural relations between galaxy scale
  radius, scale height, and stellar mass (Figs.~\ref{fig:sizes} and~\ref{fig:z50}). 

\item Gravitational scattering transfers energy from DM halo particles to stellar particles (when $\mu\geq 1$),
  causing galaxies to heat-up and expand, while haloes cool-down and contract (Fig.~\ref{fig:Vc}). As a result,
  galaxies in \HR\, are more baryon dominated than those in \LR; the latter have higher DM fractions in their inner
  regions. This result opposes the naive expectations of virtually all models for the baryon-induced contraction
  of DM haloes, and has important implications for modelling the DM fractions of galaxies using cosmological
  simulations.
  
\item Fortunately, the structural and kinematic properties of galaxies at various characteristic radii
  are robust to spurious heating provided they occupy haloes above a resolution-dependent virial mass, $M_{200}^{\rm spur}$.
  For the 100 cubic Mpc flagship run of the \eagle\, project (and our \LR\, run, both of which have 
  $m_{\rm DM}=9.7\times 10^6{\rm M_\odot}$), and for quantities measured at the stellar half-mass radius, $r_{50}$,
  this corresponds to $M_{200}\gtrsim M_{200}^{\rm spur}=10^{11.7}{\rm M_\odot}$ (or
  $N_{200}\gtrsim 5.1\times 10^4$). This coincides with the halo mass above which the {\em intrinsic} velocity
  dispersion of central galaxies in \LR\, (i.e. the values of $\sigma_{\star,50}$ measured directly from the simulated particle
  data) exceeds the {\em spurious} velocity dispersion predicted by the empirical disc heating
  model of \citet[][see Fig.~\ref{fig:sigm200}]{Ludlow2021}. Equation~(\ref{eq:exp}) therefore provides a
  simple but accurate convergence criterion that
  can be used to identify and eliminate galaxies whose structure and kinematics are influenced by numerical heating (see
  Section~\ref{ssec:conv} for details). Suppressing spurious heating at smaller/larger radii than $r_{50}$, or
  for old/young stellar populations, requires haloes to be resolved with more/fewer DM particles (see Appendix~\ref{appB}).

\end{enumerate}

Our study highlights the potential harms of spurious collisional heating for
the results of cosmological simulations of galaxy formation, and provides a simple but effective method
to identify galaxies whose structural and kinematic properties are most likely to be at risk.
Based on our analysis, we expect that galaxies in the 100 cubic Mpc-volume simulation of the \eagle\, project
\citep[][which has a DM particle mass of $m_{\rm DM}=9.7\times 10^6{\rm M_\odot}$]{Schaye2015,Crain2015} 
occupying DM haloes with masses less than $M_{200}^{\rm spur}\lesssim 10^{11.7}{\rm M_\odot}$
(or $M_\star\lesssim 10^{10}{\rm M_\odot}$) are negatively affected by spurious heating, and should not
be considered in future analyses of galaxy structure or kinematics.

We can use our model to make predictions for the values of $M_{200}^{\rm spur}$ relevant for other cosmological
simulations. For our \HR\, run, we predict spurious collisional heating to be
negligible at $r_{50}$ for galaxies occupying haloes with $M_{200}\gtrsim M_{200}^{\rm spur}=10^{11.2}{\rm M_\odot}$
(corresponding to $N_{200}\gtrsim 1.2\times 10^5$). For the 25 cubic Mpc high-resolution run of the \eagle\, project (for which
$m_{\rm DM}=1.21\times 10^6{\rm M_\odot}$) we find $M_{200}^{\rm spur}\approx 10^{11.0}{\rm M_\odot}$
(or $N_{200}^{\rm spur}\approx 8.2\times 10^4$). Similarly, for IllustrisTNG-100 we find
$M_{200}^{\rm spur}\approx 10^{11.7}{\rm M_\odot}$ ($N_{200}^{\rm spur}\approx 9.9\times 10^4$), and for
IllustrisTNG-50 \citep{Pillepich2019} we obtain\footnote{To obtain these values, we adopt a size-mass relation
appropriate for well-resolved galaxies in IllustrisTNG; specifically, $r_{50}=0.15\times r_{\rm s}$.}
$M_{200}^{\rm spur}\approx 10^{10.9}{\rm M_\odot}$ ($N_{200}^{\rm spur}\approx 2.6\times 10^5$). Note that,
to within a factor of a few, and for this specific set of cosmological simulations,
$N_{200}^{\rm spur}\sim 10^5$ (although different values of $N_{200}^{\rm spur}$ will be required to suppress
spurious heating at other galacto-centric radii). 
  
Table~\ref{TabMspur} lists these values of $M_{200}^{\rm spur}$, as well as a few additional values
for our \HR\, run that assume different relative numbers of DM and baryonic particles
(i.e. different DM particle masses at fixed baryonic particle mass). 
Note that the values of $M_{200}^{\rm spur}$ obtained for \eagle\, and IllustrisTNG exceed the lower limits on
halo masses (or the corresponding limits on stellar masses) often adopted for the analysis of these runs.
  
Although our analysis focused mainly on the stellar component of central galaxies at $z=0$, it can be 
extended to explore the redshift dependence of spurious heating, its impact on the kinematics and structure of
satellite galaxies, its effect on galaxy morphology or on the radial profiles of stellar
velocity dispersion or circular velocity. Spurious heating will also affect the fine structure of
simulated galaxies, including the abundance of bars, spiral arms, and stellar age or metallicity gradients.
We plan to investigate these topics in future work. 

Future cosmological simulations of galaxy formation would benefit from being run with
higher resolution in the DM component than what is traditionally used. The optimal value of $\mu$
is likely to be decided by separate considerations about baryon and DM mass resolutions, where $m_{\rm gas}$ is chosen
so that the sub-grid recipes for star formation and feedback yield sensible galaxy populations and $m_{\rm DM}$
is chosen so that the rate of spurious heating is reduced to tolerable levels.
The additional computational cost for the latter alone will be less than that required to increase the DM and
baryonic mass resolutions together. This will smooth the DM potential on small scales, thereby
reducing the impact of spurious heating and allowing galaxies with lower
baryonic masses to be resolved. Because its effects are primarily restricted to low masses,
the impact of spurious heating on galaxies in existing or future large-volume simulations can be quantified using 
smaller volume simulations that vary $\mu$ but not the baryonic particle mass or subgrid model parameters.

Our simulations adopted the same softening length for both resolutions (and for baryonic and DM particles).
The softening length is known to modify the effects of spurious collisional heating, suppressing it for
large values, but enhancing it for small values \citep{Ludlow2020}.
Future simulations may also profit from being carried out with optimally-chosen or adaptive softening lengths
for collisionless particles \citep[e.g.][]{Hopkins2023} to further suppress the insidious effects of
spurious collisional heating.

We hope our work stimulates others to reconsider the validity of results from existing simulations and to
make deliberate efforts to improve the reliability of future simulations.

\section*{Acknowledgements}
We thank Adrian Jenkins and Sylvia Ploeckinger for helpful conversations.
ADL and DO acknowledge financial support from the Australian Research Council through their Future Fellowship
scheme (project numbers FT160100250, FT190100083, respectively). This work benefited from the following public
{\textsc{python}} packages: {\textsc{scipy}} \citep{scipy}, {\textsc{numpy}} \citep{numpy}, {\textsc{matplotlib}}
\citep{matplotlib} and \textsc{ipython} \citep{ipython}.

\section*{Data Availability}
Our simulation data can be made available upon reasonable request. Our Theoretical results can be reproduced from the
equations provided in Section~\ref{ssec:conv}. A python script to calculate the spurious disc heating rate based on
equation~(\ref{eq:exp}) can be found at ~\url{https://github.com/AaronDLudlow/hot-disc.git}.

\bibliographystyle{mnras}
\bibliography{paper} 

\appendix

\section{Mass functions of galaxies, black holes, and dark matter haloes}
\label{appA}

\begin{figure*}
  \includegraphics[width=0.9\textwidth]{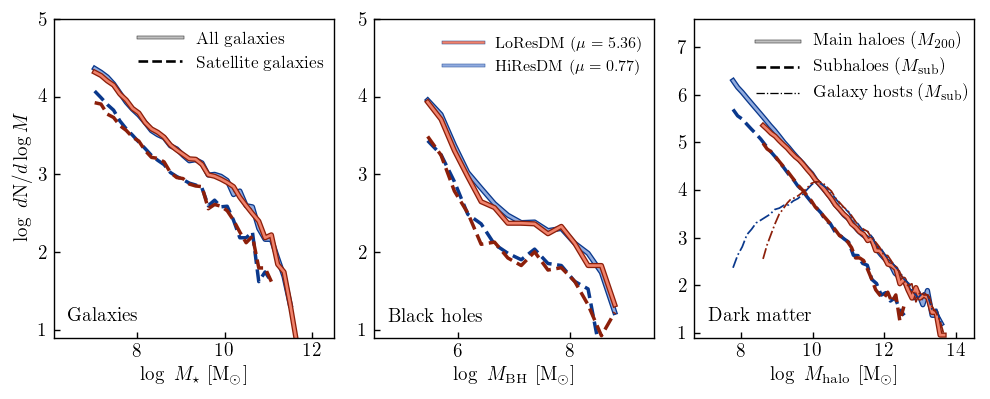}
  \caption{Mass functions for galaxies (left panel), central supermassive black holes (middle panel),
    and DM haloes and subhaloes (right panel). Results from \LR\,are shown using orange lines and those
    from \HR\, using blue lines. Stellar and central black hole mass functions are plotted for the entire
    population of galaxies (centrals plus satellites; solid lines), as well as for the subset of satellite
    galaxies (dashed lines). Dark matter halo mass functions are shown for central haloes (based on
    $M_{200}$; solid lines), for the entire population of satellite subhaloes (dashed lines), and for
    the subset of (sub)haloes (which in this case includes central and satellite haloes) hosting galaxies
    with stellar masses $M_\star\geq 5\times m_{\rm gas}$ (based on $M_{\rm sub}$; dot-dashed lines). DM halo
    mass functions are plotted to a limiting mass corresponding to 50 DM particles in each run, GSMFs are
    plotted above a limiting mass of 5 primordial gas particles, and black hole mass functions are plotted
    above $M_{\rm BH}=10^{5.5}\,{\rm M_\odot}$, which is roughly twice their seed mass. }
  \label{fig:mf}
\end{figure*}

The left panel of Fig.~\ref{fig:mf} shows the galaxy stellar mass functions (GSMF) for the full galaxy population
(solid lines) and for the subset of satellite galaxies (dashed lines). Orange and blue lines correspond to \LR\,and
\HR, respectively, and extend to a limiting mass of $M_\star = 9.1\times 10^6\,{\rm M_\odot}$ (roughly 5
primordial gas particles). The GSMFs are in excellent agreement across the entire  mass range plotted (the bumps and
wiggles at high masses are due to Poisson noise). 

In the middle panel of Fig.~\ref{fig:mf}, we plot the mass functions for the central supermassive black holes of central
(solid lines) and satellite galaxies (dashed lines). Curves are plotted to $M_{\rm BH}=10^{5.5}\,{\rm M_\odot}$, 
roughly twice the black hole seed mass. The excellent agreement between the runs extends across the entire mass range.

The right-hand panel of Fig.~\ref{fig:mf} shows the DM halo and subhalo mass functions.
We plot haloes hosting central galaxies using solid lines (their masses are
defined as $M_{200}$); dashed lines show the subhalo mass functions (whose masses are $M_{\rm sub}$). The DM halo mass
functions agree remarkably well down to a halo mass equal to about 50 DM particles, which is the lowest mass plotted.

Finally, the thin dot-dashed lines in the right-hand panel of Fig.~\ref{fig:mf} show the DM halo mass functions
(obtained using $M_{\rm sub}$) for {\rm all} DM haloes hosting luminous galaxies (i.e. those with
$M_\star\geq 5\times m_{\rm gas}$). Apart from slight discrepancies at the lowest halo masses (below about
$10^9\,{\rm M_\odot}$, which roughly corresponds to haloes resolved by $\approx 100$ DM particles in \LR), the curves
are in excellent agreement. In both simulations, these low-mass haloes contain $\lesssim 1$ per cent of the total
stellar mass, and are largely inconsequential for the total stellar mass formed in either simulation.

\section{Convergence criteria for spurious collisional heating of different stellar populations and at different galacto-centric radii}
\label{appB}

The spurious disc heating rate predicted by equation~(\ref{eq:exp})
depends on the local density and velocity dispersion of DM particles, which depend strongly on galacto-centric
radius: heating rates will be higher at $r<r_{50}$, where DM densities are higher, and lower at $r>r_{50}$, where DM densities
are lower. Furthermore, spurious heating is an integrated effect and also depends on the age of the galaxy or
stellar population: older galaxies (or stellar populations) will have experienced higher levels of heating
than younger ones, even if they occupy DM haloes with the same mass and internal structure. Below we test
our convergence criterion by applying it to galaxy properties measured at different radii, or for different
mono-age stellar populations at a fixed fiducial radius. 

\begin{figure*}
  \includegraphics[width=0.9\textwidth]{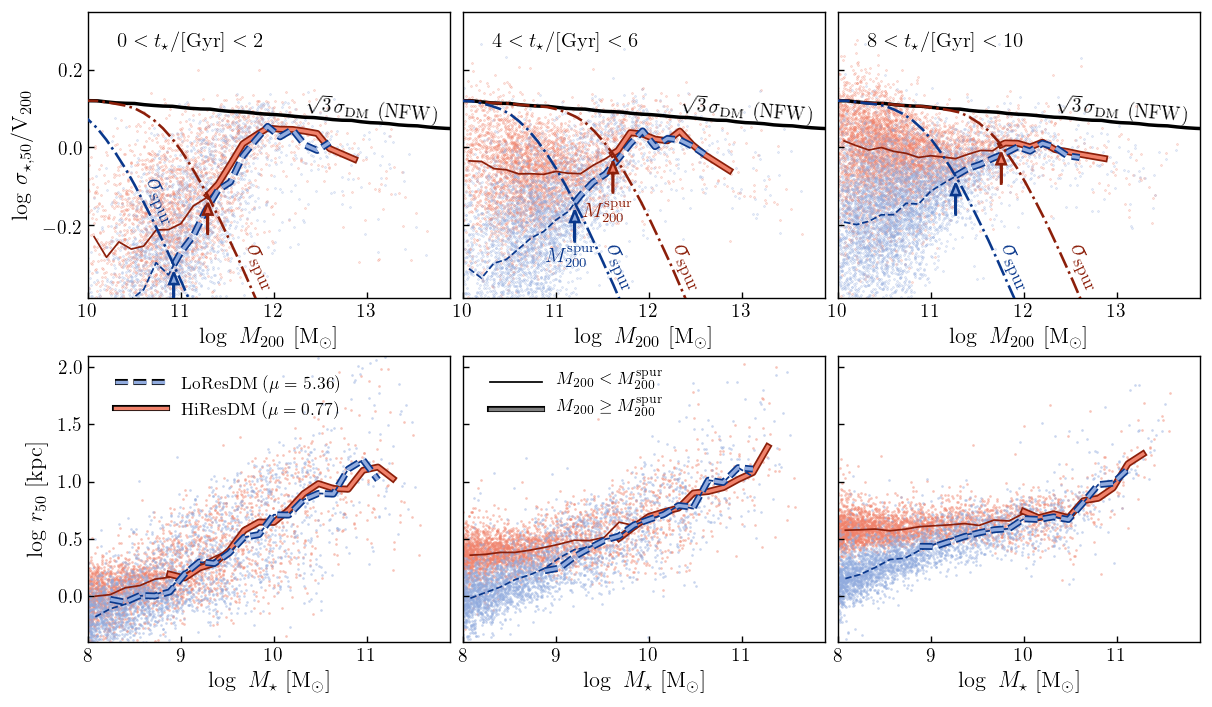}
  \caption{The importance of spurious collisional heating for mono-age stellar populations.
    The upper row is modelled after Fig.~\ref{fig:sigm200}, and plots the stellar velocity dispersion-halo mass
    relations; the lower row, similar to Fig.~\ref{fig:sizes}, plots the size-mass relations. From left to right,
    the different columns show results for stellar populations of increasing age: $0\leq t_\star/[{\rm Gyr}] \leq 2$
    (left), $4\leq t_\star/[{\rm Gyr}]\leq 6$ (middle) and $8\leq t_\star/[{\rm Gyr}] \leq 10$ (right).
    The values of $\sigma_{\star,50}$ are measured at the half-mass radius of {\em all} stars in each galaxy, not only those
    in each mono-age bin; the sizes correspond to the half stellar-mass radii of each mono-age population.
    The solid black lines in the upper panels plot the
    (isotropic) velocity dispersion of an NFW DM halo at $r_{50}$, assuming $r_{50}=0.22\times r_{\rm s}$. The dot-dashed lines
    (labelled $\sigma_{\rm spur}$ in the upper panels) show the velocity dispersion at $r_{50}$ due to spurious collisional heating
    after $t=1\,{\rm Gyr}$ (left), $5\,{\rm Gyr}$ (middle), and $9\,{\rm Gyr}$. In all panels, thin lines show the
    median results for all galaxies, whereas thick lines are limited to the subset of galaxies that pass the convergence
    criterion outlined in Section~\ref{ssec:conv}. }
  \label{fig:monoage}
\end{figure*}

\subsection{Application to mono-age stellar populations}
\label{sec:monoage}

In the upper panels of Fig.~\ref{fig:monoage}, we plot the velocity dispersion $\sigma_{\star,50}$
(normalised by $V_{200}$) as a function of $M_{200}$
for subsets of mono-age stellar particles (note: $\sigma_{\star,50}$ was measured at 
the $r_{50}$ value of {\em all} stellar particles).
Orange and blue colours distinguish our \LR\, and \HR\, runs, respectively. 
From left-to-right, the different columns correspond to
stellar populations aged between 0 to 2 Gyr, 4 to 6 Gyr, and 8 to 10 Gyr.
The solid black lines show the DM halo velocity dispersion at $r_{50}$
(where $r_{50}=0.22\times r_{\rm s}$; see Section~\ref{ssec:conv}), and the dot-dashed coloured
lines show the spurious velocity dispersion at $r_{50}$ after $t=1\,{\rm Gyr}$ (left), $3\,{\rm Gyr}$
(middle), and $9\,{\rm Gyr}$ (right) as predicted by equation~(\ref{eq:exp}).

The values of $M_{200}^{\rm spur}$ (indicated by arrows in each panel of Fig.~\ref{fig:monoage}) provide
an accurate account of the halo mass scale above which $\sigma_{\star,50}$ is unaffected by spurious heating,
for all ages of the stellar particles considered.
We show this explicitly by plotting masses $M_{200}\geq M_{200}^{\rm spur}$ using thick lines and 
$M_{200}< M_{200}^{\rm spur}$ using thin lines. 

The lower panels of Fig.~\ref{fig:monoage} confirm that our convergence criterion also applies to
the structure of mono-aged stellar particles. Here we plot (as a function of $M_\star$ rather than $M_{200}$)
the half-mass radii $r_{50}$ of each mono-aged stellar population.
Thick and thin lines differentiate the same populations as in the upper panels.

\begin{figure*}
  \includegraphics[width=0.9\textwidth]{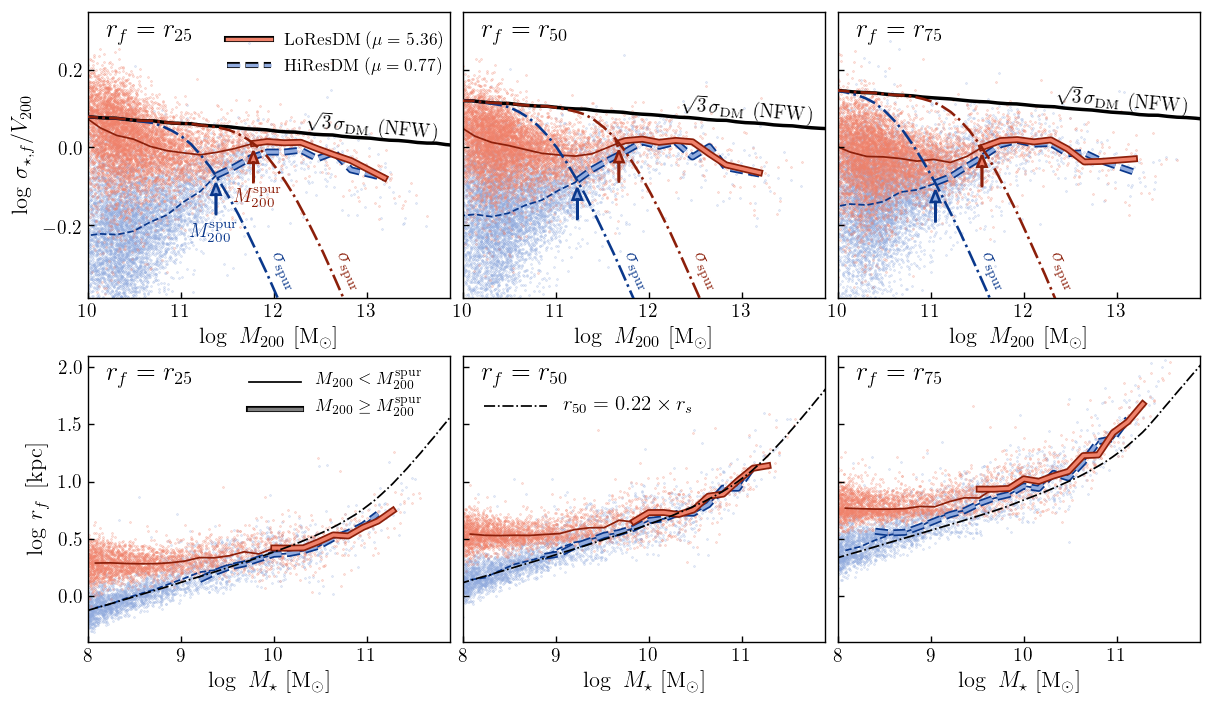}
  \caption{Same as Fig.~\ref{fig:monoage} but for the galacto-centric radii $r_{25}$ (left), $r_{50}$ (middle)
    and $r_{75}$ (right) enclosing 25, 50, and 75 per cent of the stellar mass of each galaxy.
    The dot-dashed lines in the lower-left and lower-right panels are empirical estimates of $r_{25}$
    and $r_{75}$ obtained assuming $r_{50}=0.22\times r_{\rm s}$ and a spherical exponential stellar mass profile.}
  \label{fig:spurious_rf}
\end{figure*}

\subsection{Application to different galacto-centric radii}
\label{sec:radii}

In the upper panels of Fig.~\ref{fig:spurious_rf} we plot the stellar velocity dispersion $\sigma_{\star,f}$
(normalised by $V_{200}$) at $R=r_{25}$ (left; enclosing 25 per cent of the stellar mass), $R=r_{50}$ (middle), and 
$R=r_{75}$ (right; enclosing 75 per cent of the stellar mass) as a function of $M_{200}$. The solid black lines
in each panel show the DM velocity dispersion at at these radii, assuming an isotropic NFW halo. For
haloes, we assume $r_{50}=0.22\times r_{\rm s}$ ($r_{\rm s}$ is prescribed by the mass-concentration relation of
\citealt{Ludlow2016}), and infer $r_{25}$ and $r_{75}$ assuming a spherical exponential
profile and our best-fitting stellar-to-halo mass relation in double power-law form (see footnote~\ref{fn_fit} 
and Fig.~\ref{fig:ms_mbh_mh}). The values obtained this way
roughly reproduce the median values of $r_{25}$ and $r_{75}$ measured in our simulations, as seen in the
lower panels. There we plot, in the same order, the values of $r_{25}$, $r_{50}$, and $r_{75}$ as a function of
$M_\star$, using coloured curves (and points) for the simulation results and dot-dashed black lines for the inferred
$r_f$ values. 

As in Fig.~\ref{fig:monoage}, the dot-dashed coloured lines in the upper panels (labelled $\sigma_{\rm spur}$) show
the spurious velocity dispersion at $r_f$ predicted by equation~(\ref{eq:exp}) assuming a galaxy age of
$t=7.6\,{\rm Gyr}$. The thick lines in the upper panels (solid for \LR; dashed for \HR) show the median
$\sigma_{\star,f} - {\rm M_\star}$ relations for galaxies with haloes of mass $M_{200}\geq M_{200}^{\rm spur}$,
and are in good agreement; the thin lines of each type, which differ significantly, correspond to galaxies in
lower-mass haloes. The same subsets of galaxies are distinguished in the lower panels of Fig.~\ref{fig:spurious_rf}
(i.e. using thick lines for those with $M_{200}\geq M_{200}^{\rm spur}$ and thin lines for the rest), and show that
all three characteristic sizes $r_{25}$, $r_{50}$, and $r_{75}$, as well as the velocity dispersions at these radii,
are robust to spurious collisional heating provided $M_{200}\geq M_{200}^{\rm spur}$.

\bsp	
\label{lastpage}
\end{document}